\documentclass[
reprint,
longbibliography,
superscriptaddress,
preprintnumbers,
nobibnotes,
amsmath,amssymb,
aps,
pra,
floatfix
]{revtex4-1}

\usepackage{graphicx}
\usepackage{dcolumn}
\usepackage{bm}
\usepackage{floatrow}
\usepackage[%
  colorlinks=true,
  urlcolor=blue,
  linkcolor=blue,
  citecolor=blue
]{hyperref}
\usepackage{xcolor}
\usepackage{textcomp}
\usepackage{amsmath}
\usepackage{siunitx}
\usepackage[artemisia]{textgreek}
\usepackage{upgreek}
\usepackage{etoolbox}
\usepackage[version=4]{mhchem} %chemical formulas

\usepackage{subfiles}

\begin{document}
\title{Demonstration of diamond nuclear spin gyroscope}
    
    \author{A.~Jarmola}
  \email{jarmola@berkeley.edu}
    \affiliation{
     Department of Physics, University of California,
     Berkeley, California 94720, USA
     }
      \affiliation{
    U.S. Army Research Laboratory, Adelphi, Maryland 20783, USA 
    }    
    
    \author{S.~Lourette}
    \affiliation{
     Department of Physics, University of California,
     Berkeley, California 94720, USA
     }
      \affiliation{
    U.S. Army Research Laboratory, Adelphi, Maryland 20783, USA 
    }  

\author{V.~M.~Acosta}
    \affiliation{
    Center for High Technology Materials and Department of Physics and Astronomy,
University of New Mexico, Albuquerque, New Mexico 87106, USA 
    }  
    
        \author{A.~G.~Birdwell}
    \affiliation{
    U.S. Army Research Laboratory, Adelphi, Maryland 20783, USA 
    }     
    
\author{P.~Blümler}
\affiliation{Johannes Gutenberg-Universit{\"a}t Mainz, 55128 Mainz, Germany}
    
        \author{D.~Budker}
    \affiliation{
     Department of Physics, University of California,
     Berkeley, California 94720, USA
     }
\affiliation{Johannes Gutenberg-Universit{\"a}t Mainz, 55128 Mainz, Germany}
 \affiliation{Helmholtz-Institut, GSI Helmholtzzentrum f{\"u}r Schwerionenforschung, 55128 Mainz, Germany}

    \author{T.~Ivanov}
    \affiliation{
    U.S. Army Research Laboratory, Adelphi, Maryland 20783, USA 
    }

 \author{V.~S.~Malinovsky}
    \affiliation{
    U.S. Army Research Laboratory, Adelphi, Maryland 20783, USA 
    }

\date{\today}

\begin{abstract}

We demonstrate operation of a rotation sensor based on the $^{14}$N nuclear spins intrinsic to nitrogen-vacancy (NV) color centers in diamond. The sensor employs optical polarization and readout of the nuclei and a radio-frequency double-quantum pulse protocol that monitors $^{14}$N nuclear spin precession. This measurement protocol suppresses the sensitivity to temperature variations in the $^{14}$N quadrupole splitting, and it does not require microwave pulses resonant with the NV electron spin transitions. The device was tested on a rotation platform and demonstrated a sensitivity of 4.7\,$^{\circ}/\sqrt{\rm{s}}$ (13\,mHz$/\sqrt{\rm{Hz}}$), with bias stability of 0.4\,$^{\circ}$/s (1.1\,mHz).

\end{abstract}

\maketitle

\section{Introduction}

Rotation sensors (gyroscopes) are  broadly used  for  navigation, automotive guidance, robotics, platform  stabilization, etc. \cite{VIT2017}. Various rotation-sensors technology can be subdivided into commercial and emerging. Commercial sensors include mechanical gyroscopes, Sagnac-effect optical gyroscopes, and micro-electro-mechanical systems (MEMS) gyroscopes. 
Among emerging technologies are nuclear magnetic resonance (NMR) gyroscopes. These sensors use hyperpolarized noble gas nuclei confined in vapor cells \cite{KOR2005, DON2013, Walker2016, NMRgyro2018, NMRgyro2019, SORENSEN2020, KIT2018}; they may surpass commercial devices within the next decade in terms of accuracy, robustness, and miniaturization \cite{SHE2020}.

Recently proposed nuclear spin gyroscopes based on nitrogen-vacancy (NV) color centers in diamond \cite{LED2012, AJO2012} are analogs of the vapor-based NMR devices, constituting a scalable and miniaturizable solid-state platform, capable of operation in a broad range of environmental conditions.  
An attractive feature is that a diamond sensor can be configured as a multisensor reporting on magnetic field, temperature, and strain while also serving as a frequency reference \cite{HOD2013, FAN2013, DEG2017}. This multisensing capability is important for operation in challenging environments \cite{Fu2020_Challenging}. 

In this work, we demonstrate a diamond NMR gyroscope using the $^{14}$N nuclear spins intrinsic to NV centers. Its operation is enabled by direct optical polarization and readout of the $^{14}$N nuclear spins \cite{Jarmola2020Robust} and a radio-frequency double-quantum pulse protocol that monitors $^{14}$N nuclear spin precession. This measurement technique is immune to temperature-induced variations in the $^{14}$N quadrupole splitting. In contrast to recent work \cite{JAS2019,SOS2020}, our technique does not require microwave pulses resonant with the NV electron-spin transitions. The advantage of this technique is that it directly provides information about the nuclear spin states without requiring precise knowledge of the electron spin-transition frequencies, which are susceptible to environmental influences \cite{Jarmola2020Robust}. The nuclear-spin interferometric technique developed in this work may find application in solid-state frequency references \cite{HOD2013, kraus2014} and in extending tests of fundamental interactions \cite{KOR2005, BUL2013} at the micro and nanoscale \cite{Rong2018, DING2020} to those involving nuclear spins. With further improvements, it may also find use in practical devices such as miniature diamond gyroscopes for navigational applications. 

\section{\label{sec:Results}Results}

\subsection{Experimental setup}

\begin{figure*}
\centering
    \includegraphics[width=0.9\columnwidth]{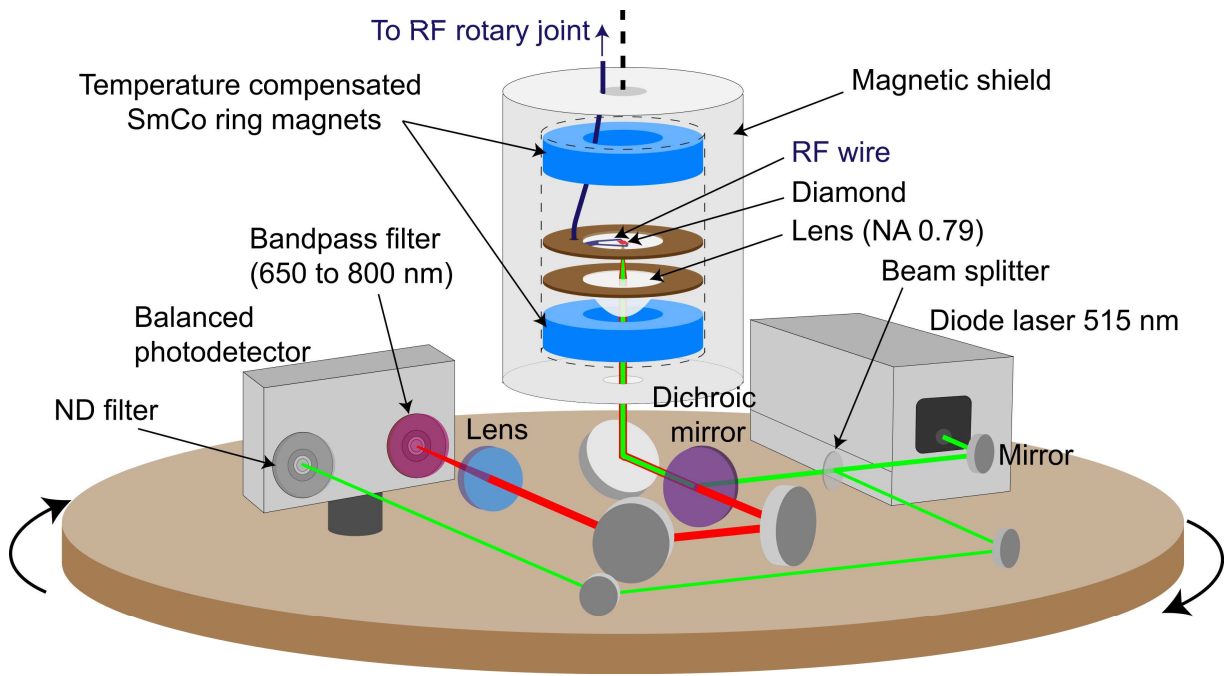}
    \caption{\label{fig:ExpSetup} \textbf{Experimental setup.} 
    Electrical connections for the laser and the photodetector are wired to the rotating platform through slip-ring lines. RF signals are delivered to the platform via single-channel RF rotary joint. ND - neutral density.
    }
\end{figure*}

Figure~\ref{fig:ExpSetup} shows an experimental setup schematic for demonstration of the diamond gyroscope. The diamond sensor, green diode laser, photodetector and all optical components are mounted on a 14-inch diameter rotating platform controlled by a commercial rate table system (Ideal Aerosmith 1291BL). 
The diamond is a 400\,$\upmu$m thick single-crystal isotopically purified (99.99\,\% $^{12}$C) diamond plate ([111] polished faces) with NV concentration of $\sim4$\,ppm. It is placed in an axial magnetic bias field of 482\,G aligned along the NV symmetry axis, providing optimal conditions for optical polarization and readout of $^{14}$N nuclear spins \cite{Jarmola2020Robust}. The bias magnetic field is produced by two temperature compensated samarium-cobalt (SmCo) ring magnets ($<\,$10\,ppm/$^{\circ}$C). These magnets were designed and arranged in a configuration that minimizes magnetic field gradients across the sensing volume. A 0.79–numerical aperture aspheric condenser lens is used to illuminate a $\sim$50-$\upmu$m-diameter spot on the diamond with 80 mW of green (515\,nm or 532\,nm) laser light and collect NV fluorescence. The fluorescence is spectrally filtered with band pass filter (650 to 800 nm) and focused onto one of the channels of a balanced photodetector. A small portion of laser light is picked off from the excitation path and directed to the other photodetector channel for balanced detection. Radio-frequency (RF) pulses for nuclear spin control are delivered using a 160\,$\upmu$m diameter copper wire placed on the diamond surface next to the optical focus. To reduce the ambient magnetic field noise, part of the setup including the diamond and magnets was placed inside low-carbon-steel magnetic shields (see sections \ref{sec:SI_Methods} and \ref{sec:SII_Magnet_Shielding}).

\begin{figure*}
\centering
    \includegraphics[width=1\textwidth]{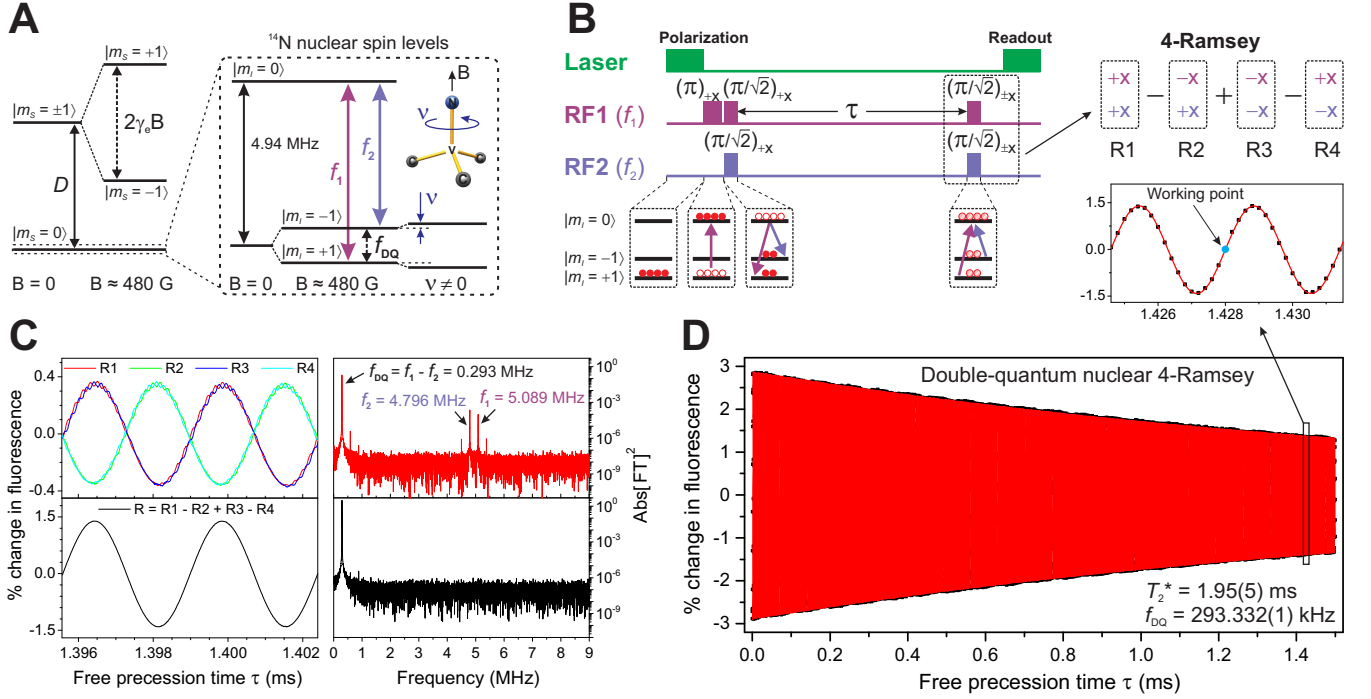}
    \caption{\label{fig:DQ} \textbf{Double-quantum measurement protocol using $^{14}$N nuclear spins in diamond}. (\textbf{A}) Energy-level diagram of the NV center ground state with and without a magnetic field $\textit{B}$ applied along the NV axis. The inset depicts the $^{14}$N nuclear spin levels, where the splitting between the $|m_{I}=\pm1\rangle$ sublevels depends on the applied field and on the rotation of the sample around the NV axis. Rotation sensing is based on the measurement of this interval. (\textbf{B}) Double-quantum nuclear Ramsey pulse sequence. (inset: 4-Ramsey phase cycling measurement scheme). (\textbf{C}) DQ nuclear Ramsey fringes (R1, R2, R3, and R4) obtained by sequentially alternating phases of the last double quantum pulse as depicted in the inset of (\textbf{B}). The frequency-domain spectrum shows the square of the absolute value of the Fourier transform, which reveals DQ signal at $f_{DQ}$ frequency and residual single-quantum signals at $f_{1}$ and $f_{2}$ frequencies. (bottom left) DQ nuclear 4-Ramsey fringes obtained by combining four DQ sequential Ramsey measurements R=R1\textminus R2+R3\textminus R4 to cancel residual SQ signals. (\textbf{D}) DQ nuclear 4-Ramsey fringes. Symbols represent experimental data, while the solid line is an exponentially decaying sine wave fit. The oscillation frequency of the signal corresponds to the $f_{DQ}$, while an exponential decay time corresponds to the nuclear DQ spin coherence time $T^{*}_{2}$= 1.95(5)\,ms. Inset: zoom to DQ 4-Ramsey fringes near the working point; rotation measurements were performed at a fixed free precession time $\tau\approx1.4$\,ms by monitoring changes in the fluorescence signal. 
    }
\end{figure*}

\subsection{Rotation detection principle}

Rotation was detected by measuring the shift in the precession rate of $^{14}$N nuclear spins intrinsic to NV centers in diamond. Figure~\ref{fig:DQ}A shows the ground state energy level diagram of the NV center. The inset shows the $^{14}$N nuclear spin states $|m_{I}\rangle$ within the $|m_{S}=0\rangle$ electron-spin state manifold. The $|m_{I}=\pm1\rangle$ states are degenerate at zero magnetic field and separated from $|m_{I}=0\rangle$ by $\sim4.94$\,MHz due to the nuclear quadrupole interaction. In the presence of an applied magnetic field $B$, the $|m_{I}=\pm1\rangle$ states experience a Zeeman shift, and the splitting frequency $f_{DQ}$ between $|m_{I}=+1\rangle$ and $|m_{I}=-1\rangle$ is described by the following equation \cite{Jarmola2020Robust}:

\begin{equation}
\begin{aligned}
\label{eq:f1-f2}
 f_{DQ}=2B\gamma_{n}\left (1-\frac{\gamma_{e}}{\gamma_{n}}\frac{A^{2}_{\perp}}{D^{2}-\gamma^{2}_{e}B^{2}}\right ),
 \end{aligned}
\end{equation} 
where $\gamma_{e}$ and $\gamma_{n}$ are the electron and nuclear $^{14}$N gyromagnetic ratios,  respectively,  $D$ is the NV electron-spin zero-field splitting parameter, and $A_{\perp}$ is the transverse magnetic hyperfine constant.

The $^{14}$N nuclear spins are prepared in a $|m_{I}=\pm1\rangle$ superposition state and precess about their quantization axis with nuclear precession frequency $f_{DQ}$.
If the diamond rotates about this axis with a frequency $\nu$, the nuclear precession frequency in the diamond reference frame is $f_{DQ}+2\nu$ (see Fig.~\ref{fig:DQ}A), where the factor of 2 arises from the fact that each state $|m_{I}=\pm1\rangle$ is shifted by $\pm\nu$. The rotation detection presented in this work is achieved by sensitively measuring these frequency shifts using a Ramsey interferometry technique.

\subsection{Measurement protocol}
Figure ~\ref{fig:DQ}B shows a double-quantum (DQ) nuclear Ramsey pulse sequence. A green laser pulse (300\,$\upmu$s duration) initializes the $^{14}$N nuclear spins into $|m_{I}=+1\rangle$ \cite{JAC2009, SME2009, STE2010PRB, FIS2013PRB}. Subsequently, an RF $\pi$ pulse (frequency $f_{1}$) is applied to transfer the population into $|m_{I}=0\rangle$. Next, to induce a double-quantum coherence between $|m_{I}=+1\rangle$ and $|m_{I}=-1\rangle$, the $f_{1}$ and $f_{2}$ transitions are simultaneously irradiated with a two-tone $\pi/\sqrt{2}$ resonant RF pulse. Then, after a free precession interval $\tau$, a second two-tone resonant RF pulse is applied to project the phase between the two nuclear spin states $|m_{I}=\pm1\rangle$ into a measurable population difference, which is read out optically. The DQ nuclear Ramsey sequence leads to population oscillations as a function of $\tau$ with a frequency $f_{DQ}+2\nu$. 

RF pulse imperfections, which arise from RF gradients across the sensing volume, lead to the appearance of residual single-quantum (SQ) signals at frequencies $f_1$ and $f_2$. These SQ transitions are sensitive to temperature fluctuations, and their signals limit the performance and robustness of the technique. To remove these SQ signals, we implemented a ``4-Ramsey'' measurement protocol, which was recently demonstrated for the NV electron spins in \cite{Hart2020}; in this work, we extend it to nuclear spins. 

The DQ nuclear 4-Ramsey protocol consists of four successive DQ Ramsey sequences with the phase of the second DQ pulse alternated (see Fig.~\ref{fig:DQ}B inset). Figure~\ref{fig:DQ}C (top left) demonstrates four optically detected $^{14}$N DQ nuclear Ramsey signals (R1, R2, R3, and R4) measured by sequentially alternating the phase of the second DQ pulse at each $\tau$ delay. The Fourier transform of the individual DQ Ramsey signals reveal residual SQ signals at $f_{1}$ and $f_{2}$ frequencies along with DQ signal at $f_{DQ}$ frequency (top right). Residual SQ signals can be eliminated by combining four DQ Ramsey signals in the following way R = R1\textminus R2+R3\textminus R4 (bottom left). The Fourier transform of the combined DQ 4-Ramsey signal shows no evidence of residual SQ signal (bottom right).

Figure~\ref{fig:DQ}D shows the optically detected $^{14}$N DQ nuclear 4-Ramsey fringes. Experimental data (symbols) are fit to an exponentially decaying sinusoidal function R$(\tau)=A \, e^{-\tau/T_{2}^{*}}  \sin(2\pi f_{DQ}\tau+\phi)$ (solid line). From the fit to the Ramsey data we infer the oscillation frequency  $f_{DQ}=293.332(1)$\,kHz and nuclear spin-coherence time $T^{*}_{2}=1.95(5)$\,ms.

Rotation measurements were performed by implementing the DQ nuclear 4-Ramsey protocol at fixed delay $\tau_{wp}=1.428$\,ms (labeled as working point on Fig.~\ref{fig:DQ}D inset) to detect changes in the phase of the oscillation. The value of $\tau_{wp}$ was selected to yield the greatest sensitivity (see section \ref{sec:SIII_Sensitivity}). Changes in the DQ 4-Ramsey signal were recorded over time and converted into frequency shifts by using a calibration coefficient. 
This coefficient is determined by values of $\tau_{wp}$ and the amplitude $A$ of the DQ 4-Ramsey fringes in the vicinity of the working point according to the expression $\alpha_0 = 4 \pi \cdot \tau_{wp} \cdot A$, and was found to be $\alpha_0$ = $6.56(6)\times10^{-5}$\,\%/($^\circ$/s) (see section \ref{sec:SIV_Calibration}). The bandwidth of the diamond gyroscope is determined by the duration of the DQ 4-Ramsey sequence $T_{cycle}$, which under current experimental conditions is $1/T_{cycle} \approx 140\,\mathrm{Hz}$.

\begin{figure}
\centering
    \includegraphics[width=1\textwidth]{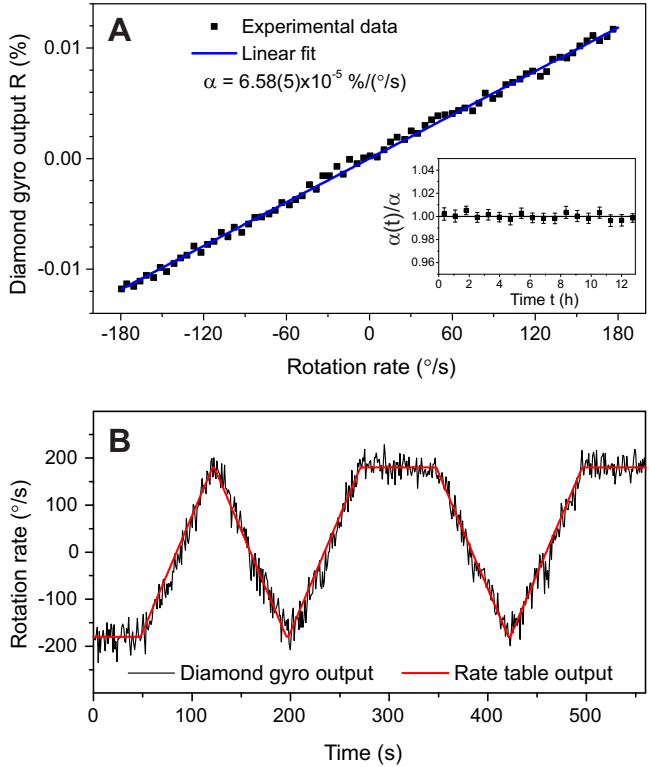}
    \caption{\label{fig:Rotation} \textbf{Diamond gyroscope: rotation sensing.} (\textbf{A}) Diamond gyroscope fluorescence signal R measured as a function of rotation rate of the platform. The calibration coefficient $\alpha$ is determined from the linear fit. \textbf{Inset}: Time-stability of $\alpha$. Fractional change in $\alpha$ is measured over several hours.
    (\textbf{B}) The rotation rates both measured by diamond gyroscope and reported by the rate table are plotted together as a function of time. The time dependence is programmed to trace ``NV''.
    }
\end{figure}

\subsection{Diamond gyroscope demonstration and noise characterization}

To demonstrate the capabilities of our diamond gyroscope across a range of rotation rates, we performed a series of test experiments on a rate table.
First we performed a calibration of the gyroscope by varying the rotation rate of the platform while measuring the DQ 4-Ramsey fluorescence signal.
This was accomplished by repeatedly sweeping the rotation rate between -180 and 180 $^\circ/s$ with a linear sweep rate of 1.8 $^\circ/s^2$ and plotting the resulting diamond gyroscope optical output signal against the rate-table rotation rate output, shown in Figure~\ref{fig:Rotation}A.
The gyroscope output exhibits a linear response over the measured range with a proportionality factor of $\alpha$ = $6.58(5)\times10^{-5}$\,\%/($^\circ$/s), which agrees with the expected value $\alpha_0$ = $6.56(6)\times10^{-5}$\,\%/($^\circ$/s) obtained from analyzing the Ramsey fringes without physically rotating the table. 

Next we used the measured value of $\alpha$ to convert the diamond gyroscope fluorescence signal into a calibrated rotation signal. Figure~\ref{fig:Rotation}B shows the rate-table output signal (red) and the diamond gyroscope signal (black), which were recorded simultaneously for various rotation sequences.

\begin{figure}
\centering
    \includegraphics[width=1\textwidth]{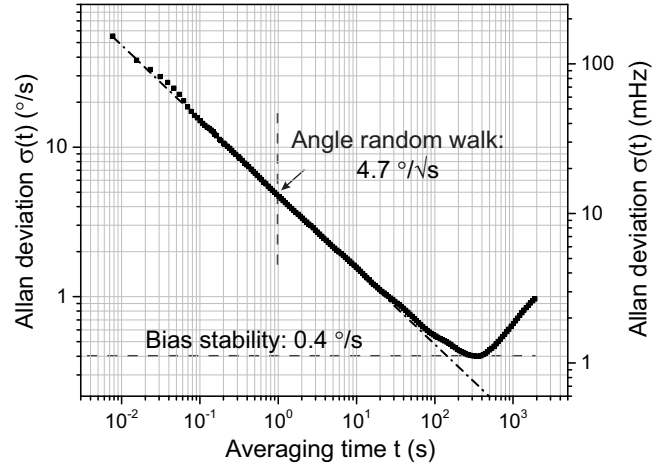}
    \caption{\label{fig:Allan} \textbf{Diamond gyroscope: Allan deviation.} Diamond gyroscope noise measurement as a function of averaging time. The diagonal dash-dotted line shows a $1/\sqrt{t}$ dependence consistent with a white frequency noise.
    }
\end{figure}

Figure~\ref{fig:Allan} shows the Allan deviation of the non-rotating diamond gyroscope noise measurement as a function of averaging time. The rotation angle random walk (ARW) sensitivity of a diamond gyroscope at current experimental conditions is 4.7\,$^\circ/\sqrt{\rm{s}}$ (13\,\rm{mHz}$/\sqrt{\rm{Hz}}$) and is near the photoelectron shot noise limit (see section SIII). The uncertainty improves until $300\,\mathrm{s}$ and reaches 0.4\,$^{\circ}$/s (1.1\,mHz) bias stability. The dashed-dotted line shows a $1/\sqrt{t}$ dependence consistent with frequency white noise.

\section{\label{sec:Discussion}Discussion and outlook}

We have demonstrated a solid-state NMR gyroscope based on $^{14}$N nuclear spins intrinsic to NV centers in diamond.
Key features of the demonstrated technique include optical polarization and readout of the nuclear spins without the use of microwave transitions and implementation of nuclear double-quantum Ramsey measurement schemes, which utilize the entire spin-1 manifold and reduce sensitivity to temperature fluctuations.
Through the use of temperature compensated magnets, magnetic shielding, and robust pulse protocols, we reduce the influence of temperature and magnetic field drift, extending the long-term stability of the gyroscope to hundreds of seconds.
The current device demonstrates a sensitivity of 4.7\,$^{\circ}/\sqrt{s}$ (13\,mHz$/\sqrt{\rm{Hz}}$) and bias stability of 0.4\,$^{\circ}$/s (1.1\,mHz). These results represent an order of magnitude improvement over those previously achieved with diamond \cite{SOS2020,JAS2019}. While the current performance of diamond gyroscope is inferior to that of vapour-based NMR gyroscopes, there is still a lot of room for improvement.

The sensitivity can be further improved by at least one order of magnitude by increasing the detected fluorescence, which can be done by increasing the laser power and by increasing the photon collection efficiency through improved collection optics and index matching \cite{SAG2012,WOL2015}.
The sensitivity can also be improved by extending the $^{14}$N nuclear spin coherence time. The present sensitivity is limited by the $^{14}$N nuclear spin $T_2^*$ which is $\sim2$\,ms. However we have measured the nuclear-spin longitudinal relaxation time $T_1$ to be $\sim80\,\rm{s}$ in the same sample. It may be possible to increase the coherence time by up to four orders of magnitude (and improve sensitivity up to two orders of magnitude) by using decoupling methods such as spin-bath driving \cite{DEL2012, BAU2018}.

The long-term stability can be improved by reducing ambient and bias magnetic field drifts. Ambient magnetic field drifts may be reduced with better magnetic shielding, while bias magnetic field drifts can be reduced by temperature stabilization of magnets. Any remaining magnetic field drifts can be corrected by employing NV-electron-spin magnetometry \cite{JAS2019}.

In future research, we will employ the $^{13}$C nuclei ensemble in diamond as an alternative sensing spin species.
At the natural abundance, these nuclei provide a higher spin density than $^{14}$N intrinsic to NV centers by several orders of magnitude. This in combination with their smaller spin-relaxation rates makes $^{13}$C a promising candidate for rotation sensing.
However, using the bulk $^{13}$C requires development of efficient methods for bulk polarization and spin readout, which is an active area of current research \cite{ALV2015, KIN2015, Ajoy2018, Pagliero2018, Sch2017, Wood2017}.

\begin{acknowledgments}
The authors are grateful to Pauli Kehayias, Daniel Thrasher, Janis Smits, Andrei Shkel, and Alexander Wood for helpful discussions. A.J. and S.L. acknowledges support from the U.S. Army Research Laboratory under Cooperative Agreement No. W911NF-18-2-0037 and No. W911NF-21-2-0030. V.M.A. acknowledges support from NSF award No. CHE-1945148, NIH award No. 1DP2GM140921, and a Cottrell Scholars award. This work was supported in part by the EU FET-OPEN Flagship Project ASTERIQS (action 820394).
\end{acknowledgments}


%Start of SUPPLEMENTAL
\clearpage
\onecolumngrid

\begin{center}
\textbf{\large Supplemental Information: Demonstration of diamond nuclear spin gyroscope}
\end{center}
\setcounter{equation}{0}
\setcounter{section}{0}
\setcounter{figure}{0}
\setcounter{table}{0}
\setcounter{page}{1}
\setcounter{equation}{0}
\setcounter{figure}{0}
\setcounter{table}{0}
\setcounter{page}{1}
\makeatletter
\renewcommand{\thetable}{S\arabic{table}}
\renewcommand{\theequation}{S\arabic{equation}}
\renewcommand{\thefigure}{S\arabic{figure}}
\renewcommand{\thesection}{S\Roman{section}}
\renewcommand{\bibnumfmt}[1]{[S#1]}
\renewcommand{\citenumfont}[1]{S#1}
\renewcommand{\thepage}{S\arabic{page}} 
%\tableofcontents

\section{\label{sec:SI_Methods}Experimental Methods}

The sample used in our experiments is a [111]-cut diamond plate (2.5\,mm\,$\times$\,1.5\,mm\,$\times$\,0.4\,mm) grown using chemical vapour deposition with an initial nitrogen concentration of $\sim13$\,ppm. NV centers were created by irradiating the sample with 4.5 MeV electrons and subsequent annealing in vacuum at 800\,$^\circ$C for one hour. The estimated NV concentration is $\sim4$\,ppm.

A linearly polarized 515-nm green diode laser light (Toptica Photonics iBeam smart 515) was used to excite NV centers in rotation experiments. Laser pulses were generated by using an external iBeam smart digital modulation option. In experiments that did not require rotation, we used 532-nm laser (Coherent Verdi G5) due to its superior noise characteristics. Laser pulses for 532-nm laser were generated by passing the continuous-wave laser beam through an acousto-optic modulator. A 0.79–numerical aperture aspheric condenser lens (Thorlabs ACL25416U-A) was used to illuminate a 50-$\mu$m-diameter spot on the diamond with $\sim80$\,mW of green light and collect fluorescence. Red fluorescence was separated from the excitation light by a dichroic mirror, passed through a 650-800\,nm bandpass filter, and detected by a balanced photodetector (Thorlabs PDB210A) producing $\sim0.1$ mA of photocurrent.

Radiofrequency pulses with carrier frequencies ($f_{1}\approx5.089$\,MHz and $f_{2}\approx4.796$\,MHz) were generated by two arbitrary waveform generators (Keysight 33512B). Each 2-channel generator was programmed to output two identical pulses on each channel that are phase shifted by 180 degrees with respect to each other. The pulses with the desired phases were subsequently selected using two TTL-controlled RF switches (Mini-Circuits ZASWA-2-50DR), combined with power combiner (Mini-Circuits ZFSC-2-6-75) and amplified by an RF amplifier (Mini-Circuits LZY-22+) (see Sec.~\ref{sec:Electronics}). RF signals were delivered to the platform via single-channel RF rotary joint.
A transistor-transistor logic (TTL) pulse card (SpinCore PBESR-PRO-500) was used to generate and synchronize the pulse sequence. A data acquisition card (National Instruments USB-6361) was used to digitize experimentally measured signals (see Sec.~\ref{sec:Readout}).

\section{\label{sec:SII_Magnet_Shielding}Permanent Magnet and Magnetic Shielding} 

\begin{figure}[b!]
    \centering
    \includegraphics[width=0.85\columnwidth]{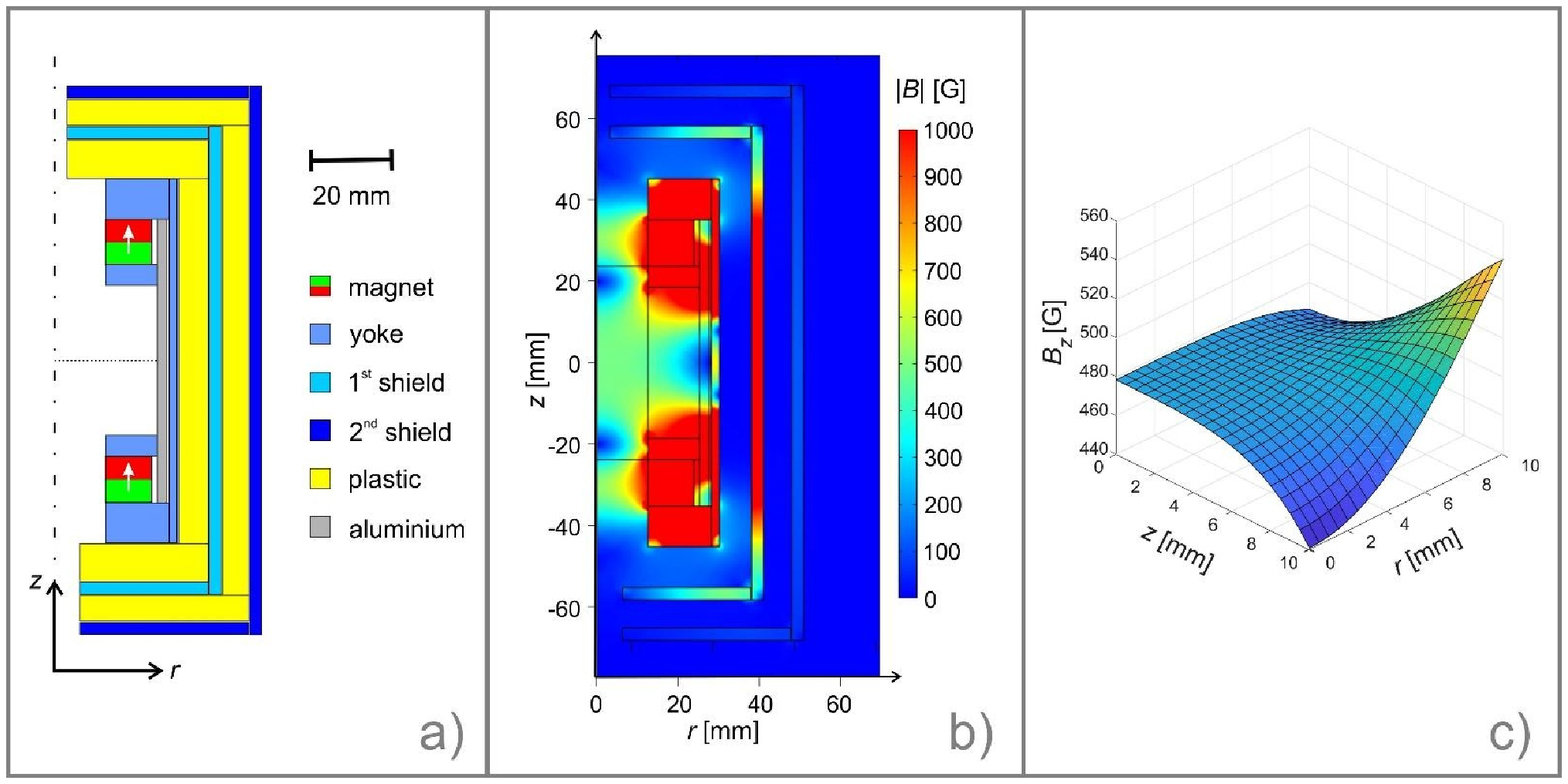}
    \caption{\textbf{Magnetic Shielding.} Axisymmetric illustrations of a) geometry and materials used for magnets, yokes and shielding. The drawing is to scale. b) Simulated flux distribution (norm of all components) inside the structure. Note that the color bar is limited for values ranging from 0 to 1000 G. c) Display of the z-flux component in the innermost region.
    }
    \label{fig:Shielding}
\end{figure}

The bias magnetic field was applied using two ring magnets made of temperature compensated \ce{Sm2Co7} (grade: EEC 2:17-TC16) with inner diameter of 1” (\SI{25.54}{mm}) and an outer diameter of 1.9” (\SI{48.26}{mm}) and a height of 0.45” (\SI{11.43}{mm}), which were purchased from Electron Energy Corporation. They possess an axial magnetization direction, an experimentally verified remanence of $B_r$ = \SI{0.794}{T}, and a temperature coefficient of $<\,$10\,ppm/$^{\circ}$C. 

The magnet-shield system was designed to meet the following requirements: a) the diamond sample should be exposed to axially directed homogeneous magnetic field with a flux density around 482 G; b) the system should shield external magnetic field, with attenuation factor on the order of $10^2$ or higher; c) there should be sufficient openings for optical access and radio-frequency cable.

The shields are made as concentric cylindrical containers of \SI{3}{\milli \meter} thickness, which significantly altered the magnetic field produced by the two ring magnets. This is because the shield acts as a yoke and redirects a significant amount of stray flux back into the center. Therefore, an additional yoke had to be made to reduce and shape the magnetic field to the desired value. Figure~\ref{fig:Shielding} shows the geometry of the magnet, yoke and shields, with the materials used indicated in the legend. 

The yoke and the shields were made from a standard steel (11SMnPb30+C with \SI{0.11}{\percent} C, \SI{0.3}{\percent} S \SIrange{0.9}{1.3}{\percent} Mn, \SIrange{0.2}{0.35}{\percent} Pb, and small amounts of Si and P, material no. 1.0718, US standard 12L13, 12L14). In the finite-element-method (FEM) simulations with COMSOL Multiphysics 5.5, the material properties of the steel parts were simulated using ``Low Carbon Steel 12L14'' from the internal COMSOL database using full B-H curves. 

The simulated attenuation of the external field was about a factor of 150; however, the experimentally determined value is about 50 (sufficient for the present experiment). We explain the discrepancy by the \SIrange{100}{200}{\micro m} gaps between the individual parts of shields that were not considered in the simulations.

When inserted into the gyro setup, the magnet and shield assembly is inverted such that the $\SI{482}{G}$ uniform magnetic field is pointed downward.

\section{\label{sec:SIII_Sensitivity}Rotation Sensitivity Calculations} 
To measure rotation rates, we prepare a superposition of a pair of nuclear spin energy eigenstates, wait for the states to acquire a relative phase over some free evolution time, and then project the accumulated phase into a population difference, which is detected with a photodiode as a reduction in fluorescence, and converted to a voltage signal. When using phase synchronized pulses, the relative phase accumulates with frequency $f$, corresponding to energy splitting. By varying the free evolution time $\tau$, we produce oscillations in the populations of the spin states.
These oscillations are detected optically and are observed as oscillations in voltage between high and low respective voltage levels, $V_H$ and $V_L$, with frequency $f$ and decay time $T_2^*$. The relationship between these quantities is described by the following equation:
\begin{align}
V(f,\tau) &= \frac{V_H + V_L}{2} \pm \frac{V_H - V_L}{2} \exp{\left(-\frac{\tau}{T_2^*}\right)} \cos{\left(2 \pi f \tau\right)}. \label{eq:Ramsey_HL}
\end{align}
The high and low voltage values can be transformed into brightness $V_0$ and contrast $C$ using the following expressions:
\begin{align}
V_0 &= \frac{V_{H} + V_{L}}{2}, & V_{H} = V_0\left(1 + \frac{C}{2}\right), \nonumber \\
C &= \frac{V_{H} - V_{L}}{V_0}, & V_{L} = V_0\left(1 - \frac{C}{2}\right) \nonumber,
\end{align}
which when applied to Eq. \ref{eq:Ramsey_HL}, gives us
\begin{align}
V(f,\tau) &= V_0\left[1 \pm \frac{C}{2} \exp{\left(-\frac{\tau}{T_2^*}\right)} \cos{\left(2 \pi f \tau\right)}\right].
\end{align}
The goal is to choose a working point $\tau = \tau_{wp}$ such that small changes in $f$ produce the largest corresponding changes in $V$, meaning that $\mathrm{d}V/\mathrm{d}f$ is maximized with respect to $\tau$. It can be demonstrated that this occurs when $\cos{\left(2 \pi f \tau_{wp}\right)} = 0$. Using this assumption, we find an expression for $\mathrm{d}V/\mathrm{d}f$ at the working point
\begin{align}
\frac{dV}{df} &= \pm \frac{V_0 \, C}{2} \exp{\left(-\frac{\tau}{T_2^*}\right)} \left(2 \pi \tau \right) \sin{\left(2 \pi f \tau\right)}, \nonumber \\
\frac{dV}{df} &= \pm \pi \, \tau \, e^{-\tau/T_2^*} \, V_0 \, C,
\end{align}
and therefore, the proportionality coefficient between $\delta f$ and $\delta V$ can be written as
\begin{equation}
    \delta f = \pm \left(\frac{1}{\pi \, \tau \, e^{-\tau/T_2^*} \, V_0 \, C}\right)\delta V. 
\end{equation}
This equation can be also written quite simply in terms of a fractional uncertainty
\begin{equation}
    \delta f = \pm \left(\frac{1}{\pi} \cdot \frac{1}{\tau \, e^{-\tau/T_2^*}} \cdot \frac{\delta V}{V_0 \, C}\right) = \pm \left(\frac{1}{\pi} \cdot \frac{1}{\tau \, e^{-\tau/T_2^*}} \cdot \frac{\delta V}{V_H - V_L}\right).
    \label{eq:sens_const}
\end{equation}

To help reduce the effect of slow drifts in laser power, instead of using the readout voltage $V$ directly, we measure and record the fractional change in fluorescence $S = V / V_{pump}$, where $V_{pump}$ is the average fluorescence signal after the spin state has been pumped back to $m_I=1$ (see \cite{Jarmola2020Robust_dup} for details). We can assume that the uncertainty of $V_{pump}$ is much smaller than that of $V$, due to a significantly longer readout time, and thus $\delta S = \delta V / V_{pump}$. Substituting these new quantities results in the following similar expressions
\begin{equation}
    \delta f = \pm \left(\frac{1}{\pi} \cdot \frac{1}{\tau \, e^{-\tau/T_2^*}} \cdot \frac{\delta S}{S_0 \, C}\right) = \pm \left(\frac{1}{\pi} \cdot \frac{1}{\tau \, e^{-\tau/T_2^*}} \cdot \frac{\delta S}{S_H - S_L}\right).
    \label{eq:sens_const2}
\end{equation}
This modification does not effect the sensitivity, because the two methods have the same fractional uncertainty $\delta S/S_0 = \delta V/V_0$.

For a photon-shot-noise-limited signal, the uncertainty in the detected signal, $\delta V_{PSN}$, is equal the detected signal divided by the square root of the number of detected photoelectrons, $\mathcal{N}_p$:
\begin{equation}
    \delta V_{PSN} = \frac{V_0}{\sqrt{\mathcal{N}_p}}. \label{eq:psn1}
\end{equation}
At higher laser powers, it is increasingly difficult to obtain a photon-shot-noise limited signal. To help reduce unwanted laser noise, the detected fluorescence signal is offset by an equivalent amount of laser light using a balanced photodiode. Doing so doubles the number of photons that contribute to photon-shot-noise, but the number of photons that contributes to the signal remains unchanged. The result is a factor of $\sqrt{2}$ increase in the uncertainty of a photon-shot-noise-limited signal
\begin{equation}
    \delta V_{PSN} = \sqrt{2} \cdot \frac{V_0}{\sqrt{\mathcal{N}_p}}. \label{eq:psn1b}
\end{equation}
The number of detected photoelectrons can be expressed as the product of the total acquisition time $T_{acq}$ and the photoelectron rate $\xi$, which in turn can be expressed as $\xi_p = V_0 / (G q_e)$, where $V_0$ is the output of the photodiode, $G$ is the gain (transimpedance) of the photodiode, and $q_e$ is the (positive) charge of an electron.
The acquisition time can be expressed as $T_{acc} = t_R \, N_{meas}$, where $t_R$ the readout time of a measurement, and $N_{meas}$ is the number of measurements. Combining these into one expression for $\mathcal{N}_p$, we have
\begin{align}
\mathcal{N}_p &= \xi_{p} \cdot T_{acc}, \nonumber \\
\mathcal{N}_p &= \left(\frac{V_0}{G \, q_e} \right) \left(t_R N_{meas} \right). \label{eq:psn2}
\end{align}
These two equations (\ref{eq:psn1b} and \ref{eq:psn2}) can be combined to obtain an expression for the uncertainty in the photon-shot-noise-limited signal
\begin{align}
\frac{\delta V_{PSN}}{V_0} &= \frac{\sqrt{2}}{\sqrt{\mathcal{N}_p}} = \sqrt{\frac{2 \, G \, q_e}{V_0 \, t_R}} \sqrt{\frac{1}{ N_{meas}}}.
\end{align}
Thus, we have obtained an expression for the photon-shot-noise-limited fractional uncertainty of the spin state in terms of known experimental conditions.
Using equation \ref{eq:sens_const} (or \ref{eq:sens_const2}), this uncertainty can be converted into a corresponding frequency measurement uncertainty
\begin{align}
\delta f_{PSN} = \left(\frac{1}{\pi} \cdot \frac{1}{\tau \, e^{-\tau/T_2^*}}\right) \cdot \frac{\delta V_{PSN}}{V_0 \, C} &= \left(\frac{1}{\pi} \cdot \frac{1}{\tau \, e^{-\tau/T_2^*}}\right) \cdot \frac{1}{C}\sqrt{\frac{2 \, G \, q_e}{V_0 \, t_R}} \sqrt{\frac{1}{N_{meas}}}\nonumber\\
\delta f_{PSN} &= \left(\frac{1}{\pi} \cdot \frac{1}{\tau \, e^{-\tau/T_2^*}} \cdot \frac{1}{C} \sqrt{\frac{2 \, G \, q_e}{V_0 \, t_R}} \right) \sqrt{\frac{t_{meas}}{t}}.
\end{align}
Here $t_{meas}$ is the time of a single measurement, and $t$ is the total measurement time. Note that it is required that $t_{meas} > \tau$.

Each of the spin states used in the Double Quantum 4-Ramsey protocol ($m_I = +1$ and $m_I = -1$) is sensitive to changes in the rotation rate, but with opposite sign. This results in rotation-induced shifts of the splitting frequency that are twice as large as the corresponding change in rotation frequency:
\begin{align}
    \Delta \nu_{DQ} &= \frac{\Delta f_{DQ}}{2}, \label{eq:FactorOfTwo}
\end{align}
which results in following expression for the photon-shot-noise-limited uncertainty in rotation rate
\begin{align}
\delta \nu_{PSN} &= \left(\frac{1}{2\pi} \cdot \frac{1}{\tau \, e^{-\tau/T_2^*}} \cdot \frac{1}{C} \sqrt{\frac{2 \, G \, q_e}{V_0 \, t_R}} \right) \sqrt{\frac{t_{meas}}{t}}.
\end{align}
The quantities $\tau$, $t_R$, and $t_{meas}$ were chosen to optimize the measured sensitivity for the experiment.
For these values, we calculate a photon-shot-noise-limited sensitivity of $\SI{9.8}{\milli\hertz\per\sqrt{\hertz}}$ for Double Quantum nuclear Ramsey experiments
\begin{align}
\delta \nu_{PSN}  &= \left(\frac{1}{2 \pi} \cdot \frac{1}{\SI{1.4}{\milli s}\cdot e^{-\SI{1.4}{ms}/\SI{2.0}{ms}}} \cdot \frac{1}{0.015} \cdot \sqrt{\frac{2 \cdot (\SI{175}{kV\per A}) \cdot (\SI{1.6e-19}{A \per \hertz})}{(\SI{15}{V}) \cdot (\SI{17}{\mu s})}} \right) \sqrt{\frac{\SI{1.92}{\milli s}}{t}} \nonumber \\
&= \left(\SI{9.8}{\milli\hertz\per\sqrt{\hertz}}\right) \cdot t^{-1/2}.
\end{align}

\section{\label{sec:SIV_Calibration}Calibration Coefficient}
The NV gyroscope fundamentally does not require calibration: shifts of the frequency of the measured Ramsey fringes (see Fig.~2D) directly reports changes of the rotation rate.
However, instead of scanning $\tau$ over the full Ramsey signal, we measured only at the fixed value $\tau$ that yields the optimal sensitivity (see Fig.~2D-inset labeled as working point and Sec.~\ref{sec:SIII_Sensitivity}).
When operating in this mode, changes in the oscillation frequency are measured by detecting a phase shift, which are then converted into a corresponding frequency shift.
In this case, one needs to determine the calibration coefficient that converts the detected signal (fractional change in fluorescence) to the rotation rate.
For sufficiently small rotation rates ($|\nu| \ll 1/\tau$), the fractional change in fluorescence is proportional to the rotation rate.
The proportionality coefficient can be measured directly (through rotation) or indirectly (by varying $\tau$) in order to calibrate the gyroscope.
    
\begin{figure}[b]
    \centering
    \includegraphics[width=1\columnwidth]{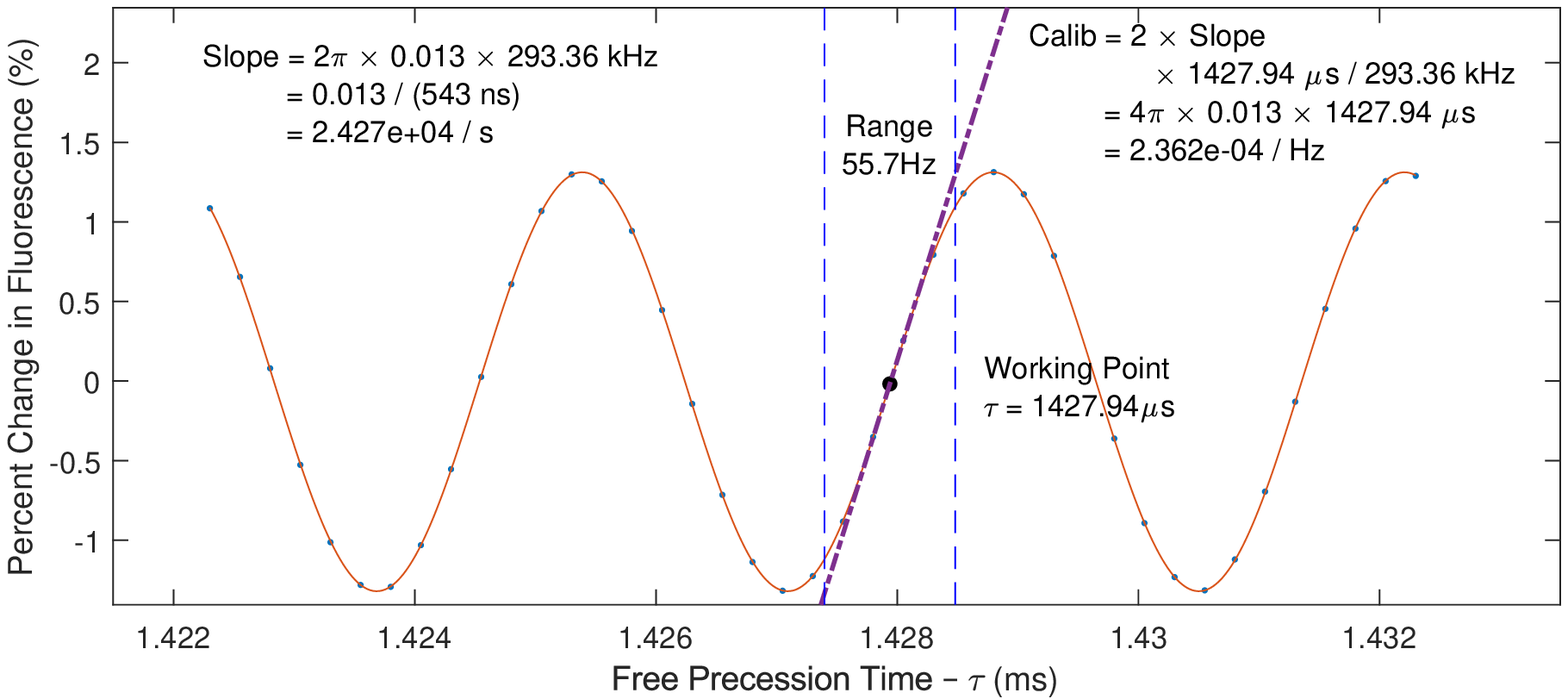}
    \caption{\textbf{Calibration Coefficient.} DQ 4-Ramsey fringes in the vicinity of the working point are fit to a sinusoid. Fit parameters are used to determine the calibration coefficient $\alpha$. 
    }
    \label{fig:CalibrationCoefficient}
\end{figure} 
    
The calibration coefficient can be readily obtained by varying $\tau$ to measure the amplitude of the Ramsey fringes in the vicinity of the working point, shown in Figure \ref{fig:CalibrationCoefficient}. The calibration coefficient, which we define to be the proportionality factor between the change in rotation rate $\Delta\nu$ and change in detected signal $\Delta S$, is described by the equation
\begin{align}
    \Delta S = \left(\frac{S_0}{\nu_0}\right) \Delta \nu = \left(4 \pi \tau \, S_0\right) \Delta \nu &= \left(\frac{\SI{1.32}{\percent}}{\SI{55.7}{\hertz}}\right) \Delta \nu \nonumber \\
    \alpha = \frac{\Delta S}{\Delta \nu} &= \SI{2.36e-2}{\percent \per Hz}  \nonumber \\
    &= \SI{6.56e-5}{\percent \per \left(^\circ\per s\right)} ,
\end{align}
where $S_0$ is the amplitude of the Ramsey fringes near the working point and $\nu_0$ is the rotation rate that shifts the phase of the fringes by $\SI{1}{\radian}$ (see Eq.~\ref{eq:1radshift}). This method produces precise estimates of $\alpha$, but is more sensitive to bias that can arise from deviations from the modelling function.

The same quantity can also be obtained through other methods. A similar method involves carefully measuring $\mathrm{d}S/\mathrm{d}\tau$ at the working point, and then converting to $\mathrm{d}S/\mathrm{d}\nu$
\begin{align}
    \alpha = \frac{\mathrm{d} S}{\mathrm{d} \nu} = 2 \cdot  \frac{\mathrm{d}S}{\mathrm{d}f_{DQ}} &= 2\left( \frac{\tau_{wp}}{f_{DQ}} \cdot \frac{\mathrm{d}S}{\mathrm{d}\tau}\right)
\end{align}
A different approach for obtaining the calibration coefficient involves applying a known magnetic field and measuring the induced shift in the output signal when measuring at the working point. All three of these methods were found to produce identical measurements of $\alpha$.

\section{Dynamic Range and Linearity} \label{sec:Dynamic Range}
To characterize the dynamic range and deviation from linearity of our gyroscope, we first define the rotation rate $\nu_0$  that corresponds to a $\SI{1}{\radian}$ shift in the phase of the Ramsey fringes, including the factor of 2 between line shift $\Delta f_{DQ}$ and rotation rate $\nu$ (see Eq. \ref{eq:FactorOfTwo}), according to the following equation
\begin{align}
\nu_0 &\equiv \frac{1}{2} \cdot\frac{1}{2 \pi \tau} = \SI{55.7}{Hz}\label{eq:1radshift}.
\end{align}
We can think of $\pm\nu_0$ as the approximate range of the gyroscope over which rotation rates can be unambiguously and precisely determined. As $\nu$ increases and becomes similar in size to $\nu_0$, the measured value of the rotation rate $\nu_{meas}$ begins to underestimate the magnitude of the rotation rate according to the formula
\begin{align}
\frac{\nu_{meas}}{\nu_0} &= \sin\left(\frac{\nu}{\nu_0}\right).
\end{align}
After Taylor expanding the sinusoid about $\nu = 0$, we obtain an expression for the fractional deviation from linearity as a function of the rotation rate $\nu$, for $\left|\nu\right| \ll \nu_0$
\begin{align}
\epsilon(\nu) \equiv \frac{\nu - \nu_{meas}}{\nu} &= \frac{1}{3!}\left(\frac{\nu}{\nu_0}\right)^2 - \, \frac{1}{5!}\left(\frac{\nu}{\nu_0}\right)^4 + \quad \ldots \, .
\end{align}
If we require a certain tolerance $\epsilon$ for the fractional deviation from linearity, we can calculate the corresponding dynamic range $-\nu_{DR} < \nu < \nu_{DR}$ of our device
\begin{align}
\epsilon &\approx \num{5.2e-5} \cdot \left(\frac{\nu_{DR}}{\SI{1}{Hz}}\right)^2 \nonumber \\
\nu_{DR} &= \SI{140}{Hz} \cdot \sqrt{\epsilon}.
\end{align}
If it is required that the gyroscope deviates from linearity by at most 100 ppm, the dynamic range of the gyroscope is $\pm \SI{1.4}{Hz}$ or $\pm\SI{500}{^\circ\per s}$. The dynamic range can be improved by choosing a smaller value for $\tau$, or made arbitrarily large through the implementation of feedback.

\begin{figure}
\centering
    \includegraphics[width=1\columnwidth]{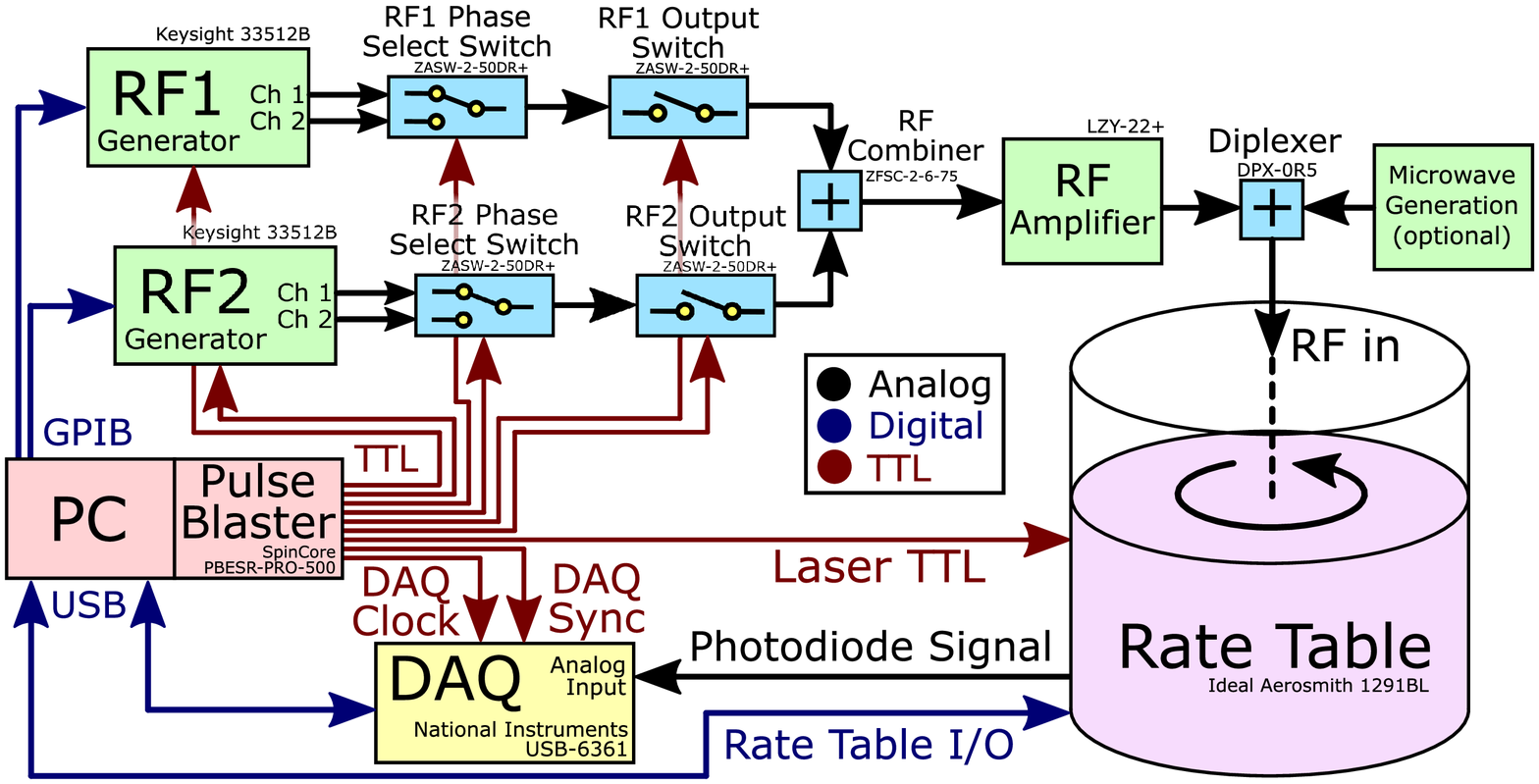}
    \caption{\textbf{Electronics Schematic.} Electronics are responsible for RF generation, laser-pulse controls, rate table I/O, and data acquisition. \textbf{TTL Pulse Generation:} TTL pulses are generated with a PulseBlaster PCI card. \textbf{RF Generation:} RF is generated with two devices, labelled RF1 and RF2, on or near frequencies $f_1$ and $f_2$, respectively. Each device outputs on two channels, 180 degrees out of phase, and is selected with an RF Phase Select Switch. RF output switches are used to gate the RF signal into pulses with desired timing and duration. The shaped pulses are combined with an RF Splitter, amplified with an RF Amplifier, and sent to the rotational platform (Rate Table) through an RF rotary joint at the top of the rotation axis. The pulseblaster provides TTL pulses for triggering the pulse generation, selecting the phase, and shaping/gating the output. \textbf{Data Acquisition:} The photodoide output signal is sent through a slip-ring line of the rotation platform and into the analog input port of the data acqusition device (DAQ). The DAQ receives two TTL signals from the pulse blaster: a DAQ sample clock, which determines when the DAQ should sample the analog input, and a DAQ Sync, to indicate the start of an experimental cycle. \textbf{Rate Table:} The controlling PC sets the rotation rate (setpoint) of the rate table, and continuously reads out the actual value of the rotation rate, using the Application Programming Interface (API) of the Rate Table. \textbf{Laser:} The PulseBlaster provides a TTL signal to the laser through a slip-ring line of the rotation platform, to generate the optical pulses. 
    }
    \label{fig:electronics}
\end{figure}

\section{Electronics} \label{sec:Electronics}

A graphical representation of the electrical connections used when performing a Double Quantum (DQ) nuclear 4-Ramsey experiment is shown in Fig.~\ref{fig:electronics}.
To perform the experiments outlined in this text, LabVIEW applications were developed for a central computer to communicate with many devices.
All of the transistor-transistor logic (TTL) signals (shown in red) that are sent by the computer are provided by the Pulse Blaster PRO, a programmable PCI card with an oven-controlled clock oscillator with 200 ppb stability, capable of producing up to 21 digital outputs with \SI{2}{ns} resolution.
At the start of an experiment, it is programmed to output cyclical pulse sequences on nine different output channels, serving as a clock for the experiment.
Although it is a PCI card, it can be thought of as an independent device: once it is programmed at the start of an experiment, it will run with no input from the computer indefinitely.

Two Keysight 33512B function generators, labelled as RF1 and RF2, are used to produce RF signals at or near frequencies $f_1$ and $f_2$, respectively. Upon receiving a TTL signal, each device outputs two otherwise identical pulses that are phase shifted by 180 degrees, and each pair of outputs is connected to an ``RF phase select'' switch, which is used to select the desired phase of the RF pulse. After selecting the phase, the signals from the two devices are passed through a second pair of ``RF output'' switches, which serve as output gates for each RF device. After both pairs of switches, the signals are combined into a single output using an RF splitter, amplified using an RF amplifier, and delivered to the diamond sample through a small loop of wire.

When operating as a gyroscope, all of the RF pulses can be programmed as an arbitrary waveform into a single channel of a single function generator. When doing so, only a single TTL pulse is required to synchronize the pulses with the optical pulses.

\section{Data Acquisition -- Optical Readout} \label{sec:Readout}

\begin{figure}[h!]
    \centering
    \includegraphics[width=1\columnwidth]{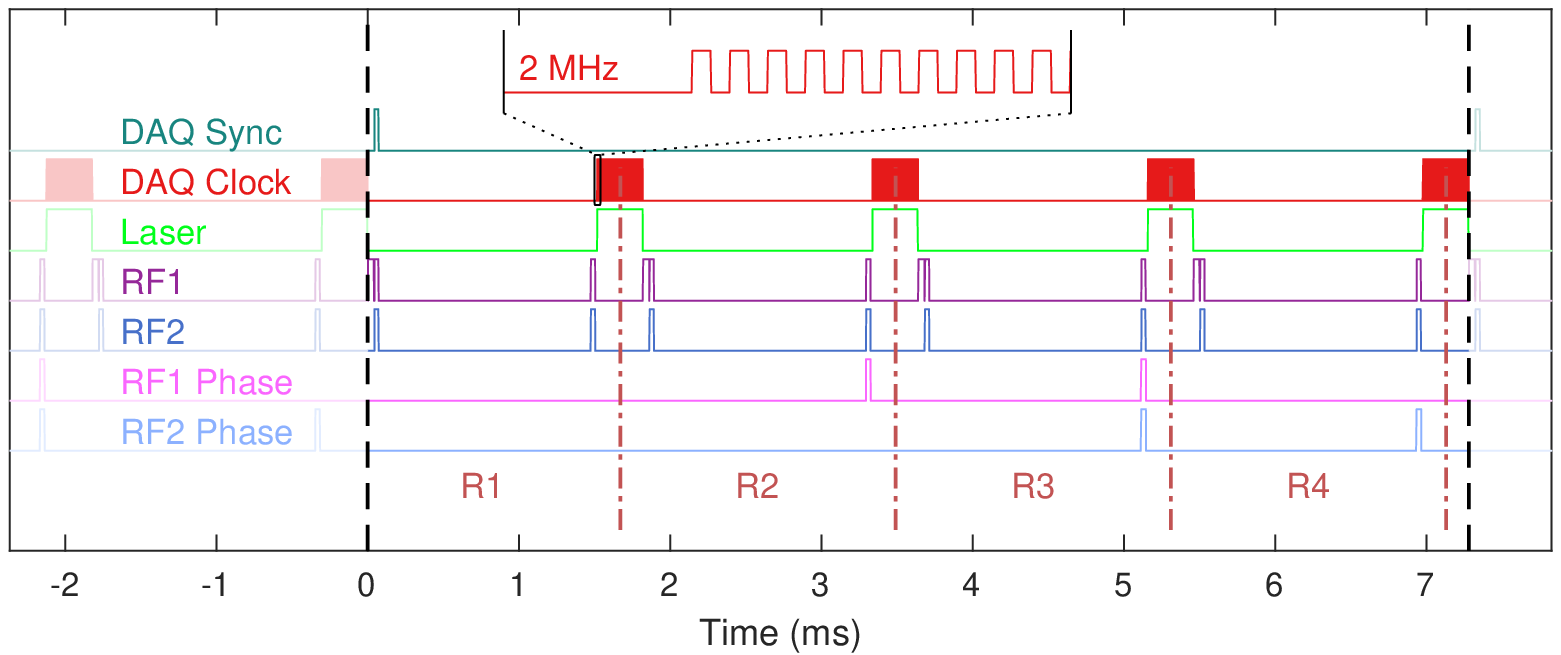}
    \caption{\textbf{Double Quantum 4-Ramsey Pulse Sequence Diagram.} Digital TTL pulses are sent from the Pulse Blaster to DAQ, laser, and RF devices. The pulse sequence can be divided into four different Ramsey experiments, denoted as R1, R2, R3, and R4. The DAQ Clock signal triggers the acquisition of of a single voltage reading. The DAQ Sync signal is used to ensure that the DAQ begins processing DAQ Clock pulses at the start of the experimental cycle. All pulses are drawn to scale.
    }
    \label{fig:DigitalWaveform}
\end{figure}

To perform the optical readout of the $^{14}$N nuclear spin state, fluorescence from the NV center is captured with a lens and focused onto a photodiode.
The photodiode used for obtaining the results reported in the main text is Thorlabs PDB210A, which is a differential photodiode.
The output of the photodiode is sent (through the rate-table port) to an analog input (AI) port of a Data Acqusition (DAQ) card (NI USB-6361). 
Additionally, the DAQ receives two digital triggers, which are provided by the PulseBlaster. The pulses used during a Double Quantum nuclear 4-Ramsey experimental sequence are shown in Fig~\ref{fig:DigitalWaveform}.

When acquiring data for an experiment, the DAQ first waits for a synchronization pulse (DAQ Sync), and then begins sampling and saving the voltage of the Analog Input (AI) to a buffer on the rising edge of each clock signal, until the buffer is filled.

For the duration of each optical readout of the NV center, the Pulse Blaster PRO provides a continuous clock signal (DAQ Clock) at \SI{2}{MHz}, corresponding to the maximum sample rate of the DAQ.
After the desired number of repeats---typically corresponding to \SI{1}{s} of data acquisition, $\sim 150$ experimental cycles, or $\sim 600$ optical pulses---the output buffer is filled and the data are sent to the computer over USB.

A typical experiment may have 600 voltage readings ($\SI{300}{\micro s} \times \SI{2}{MHz}$) per optical pulse, 4 optical pulses (R1, R2, R3, R4) per experimental cycle, and 150 experimental cycles per second (cycle duration $\SI{7}{ms}$). If a data transfer occurs once every second, then this corresponds to around $600 \times 4 \times \SI{150}{\per s} \times  \SI{4}{B} = \SI{1.5}{MB\per s}$ of binary data, significantly larger in ASCII format. Although this is not an excessive amount of data, after many hours it exceeds a typical quantity of computer RAM, making it cumbersome to analyze. 
Thus, two methods were used to reduce the quantity of data before saving to the hard drive, more for convenience than necessity. 
The first method is averaging the 100 experimental cycles into a single averaged experimental cycle. The second involves processing and extracting the contrast from each optical pulse (300 voltage readings), which is done by applying the appropriate integration windows to obtain the projected spin state of the NV center for each optical readout. In the simplest case, both methods can be performed, reducing the data to a single value for each optical readout in the experimental cycle (typically 4). Finally, to minimize downtime and streamline the acquisition process, all data processing and saving is done while the computer is waiting for the DAQ to acquire and transfer the next buffer of data.

\section{Rotation Rate Table}  \label{sec:Rate Table}
The computer uses the rate table's Application Programming Interface (API) to set its motion to a constant rotation rate using the ``JOG'' command. Subsequent issues of this command cause the table's rotation rate setpoint to linearly ramp from its prior value to the new desired velocity, using feedback to ensure that the actual velocity closely follows the setpoint. The slope of the linear ramp is the angular acceleration, which is set to the desired value of the ramp rate using the ``ACL'' command. Clockwise rotations are defined to be positive.

The rate table was controlled by a separate LabVIEW application from the application responsible for instrument control and data acquisition. This application is given list of instructions, each of which contains a timestep duration, angular velocity, and angular acceleration. For each instruction in the list, the applcation waits for the specified time, and then updates the angular velocity setpoint and angular acceleration, accordingly. The application also polls the rate table's current angle, rotation rate, and angular acceleration every \SI{30}{ms} and records these values along with the current timestamp to a file.


\begin{thebibliography}{37}%
\makeatletter
\providecommand \@ifxundefined [1]{%
 \@ifx{#1\undefined}
}%
\providecommand \@ifnum [1]{%
 \ifnum #1\expandafter \@firstoftwo
 \else \expandafter \@secondoftwo
 \fi
}%
\providecommand \@ifx [1]{%
 \ifx #1\expandafter \@firstoftwo
 \else \expandafter \@secondoftwo
 \fi
}%
\providecommand \natexlab [1]{#1}%
\providecommand \enquote  [1]{``#1''}%
\providecommand \bibnamefont  [1]{#1}%
\providecommand \bibfnamefont [1]{#1}%
\providecommand \citenamefont [1]{#1}%
\providecommand \href@noop [0]{\@secondoftwo}%
\providecommand \href [0]{\begingroup \@sanitize@url \@href}%
\providecommand \@href[1]{\@@startlink{#1}\@@href}%
\providecommand \@@href[1]{\endgroup#1\@@endlink}%
\providecommand \@sanitize@url [0]{\catcode `\\12\catcode `\$12\catcode
  `\&12\catcode `\#12\catcode `\^12\catcode `\_12\catcode `\%12\relax}%
\providecommand \@@startlink[1]{}%
\providecommand \@@endlink[0]{}%
\providecommand \url  [0]{\begingroup\@sanitize@url \@url }%
\providecommand \@url [1]{\endgroup\@href {#1}{\urlprefix }}%
\providecommand \urlprefix  [0]{URL }%
\providecommand \Eprint [0]{\href }%
\providecommand \doibase [0]{http://dx.doi.org/}%
\providecommand \selectlanguage [0]{\@gobble}%
\providecommand \bibinfo  [0]{\@secondoftwo}%
\providecommand \bibfield  [0]{\@secondoftwo}%
\providecommand \translation [1]{[#1]}%
\providecommand \BibitemOpen [0]{}%
\providecommand \bibitemStop [0]{}%
\providecommand \bibitemNoStop [0]{.\EOS\space}%
\providecommand \EOS [0]{\spacefactor3000\relax}%
\providecommand \BibitemShut  [1]{\csname bibitem#1\endcsname}%
\let\auto@bib@innerbib\@empty
%</preamble>
\bibitem [{\citenamefont {Passaro}\ \emph {et~al.}(2017)\citenamefont
  {Passaro}, \citenamefont {Cuccovillo}, \citenamefont {Vaiani}, \citenamefont
  {De~Carlo},\ and\ \citenamefont {Campanella}}]{VIT2017}%
  \BibitemOpen
  \bibfield  {author} {\bibinfo {author} {\bibfnamefont {Vittorio}\
  \bibnamefont {Passaro}}, \bibinfo {author} {\bibfnamefont {Antonello}\
  \bibnamefont {Cuccovillo}}, \bibinfo {author} {\bibfnamefont {Lorenzo}\
  \bibnamefont {Vaiani}}, \bibinfo {author} {\bibfnamefont {Martino}\
  \bibnamefont {De~Carlo}}, \ and\ \bibinfo {author} {\bibfnamefont
  {Carlo~Edoardo}\ \bibnamefont {Campanella}},\ }\bibfield  {title} {\enquote
  {\bibinfo {title} {Gyroscope technology and applications: A review in the
  industrial perspective},}\ }\href {\doibase 10.3390/s17102284} {\bibfield
  {journal} {\bibinfo  {journal} {Sensors}\ }\textbf {\bibinfo {volume} {17}},\
  \bibinfo {pages} {2284} (\bibinfo {year} {2017})}\BibitemShut {NoStop}%
\bibitem [{\citenamefont {Kornack}\ \emph {et~al.}(2005)\citenamefont
  {Kornack}, \citenamefont {Ghosh},\ and\ \citenamefont {Romalis}}]{KOR2005}%
  \BibitemOpen
  \bibfield  {author} {\bibinfo {author} {\bibfnamefont {T.~W.}\ \bibnamefont
  {Kornack}}, \bibinfo {author} {\bibfnamefont {R.~K.}\ \bibnamefont {Ghosh}},
  \ and\ \bibinfo {author} {\bibfnamefont {M.~V.}\ \bibnamefont {Romalis}},\
  }\bibfield  {title} {\enquote {\bibinfo {title} {Nuclear spin gyroscope based
  on an atomic comagnetometer},}\ }\href {\doibase
  10.1103/PhysRevLett.95.230801} {\bibfield  {journal} {\bibinfo  {journal}
  {Phys. Rev. Lett.}\ }\textbf {\bibinfo {volume} {95}},\ \bibinfo {pages}
  {230801} (\bibinfo {year} {2005})}\BibitemShut {NoStop}%
\bibitem [{\citenamefont {Donley}\ and\ \citenamefont
  {Kitching}(2013)}]{DON2013}%
  \BibitemOpen
  \bibfield  {author} {\bibinfo {author} {\bibfnamefont {E.~A.}\ \bibnamefont
  {Donley}}\ and\ \bibinfo {author} {\bibfnamefont {J.}~\bibnamefont
  {Kitching}},\ }\enquote {\bibinfo {title} {Nuclear magnetic resonance
  gyroscopes},}\ in\ \href {\doibase 10.1017/CBO9780511846380.020} {\emph
  {\bibinfo {booktitle} {Optical Magnetometry}}},\ \bibinfo {editor} {edited
  by\ \bibinfo {editor} {\bibfnamefont {Dmitry}\ \bibnamefont {Budker}}\ and\
  \bibinfo {editor} {\bibfnamefont {Derek~F.}\ \bibnamefont
  {Jackson~Kimball}}}\ (\bibinfo  {publisher} {Cambridge University Press},\
  \bibinfo {year} {2013})\ p.\ \bibinfo {pages} {369–386}\BibitemShut
  {NoStop}%
\bibitem [{\citenamefont {Walker}\ and\ \citenamefont
  {Larsen}(2016)}]{Walker2016}%
  \BibitemOpen
  \bibfield  {author} {\bibinfo {author} {\bibfnamefont {T.~G.}\ \bibnamefont
  {Walker}}\ and\ \bibinfo {author} {\bibfnamefont {M.~S.}\ \bibnamefont
  {Larsen}},\ }\enquote {\bibinfo {title} {Chapter eight - spin-exchange-pumped
  {NMR} gyros},}\ in\ \href
  {https://www.sciencedirect.com/science/article/pii/S1049250X16300064} {\emph
  {\bibinfo {booktitle} {Advances In Atomic, Molecular, and Optical
  Physics}}},\ Vol.~\bibinfo {volume} {65}\ (\bibinfo  {publisher} {Academic
  Press},\ \bibinfo {year} {2016})\ pp.\ \bibinfo {pages}
  {373--401}\BibitemShut {NoStop}%
\bibitem [{\citenamefont {Limes}\ \emph {et~al.}(2018)\citenamefont {Limes},
  \citenamefont {Sheng},\ and\ \citenamefont {Romalis}}]{NMRgyro2018}%
  \BibitemOpen
  \bibfield  {author} {\bibinfo {author} {\bibfnamefont {M.~E.}\ \bibnamefont
  {Limes}}, \bibinfo {author} {\bibfnamefont {D.}~\bibnamefont {Sheng}}, \ and\
  \bibinfo {author} {\bibfnamefont {M.~V.}\ \bibnamefont {Romalis}},\
  }\bibfield  {title} {\enquote {\bibinfo {title}
  {$^{3}\mathrm{He}\text{\ensuremath{-}}^{129}\mathrm{Xe}$ comagnetometery
  using $^{87}\mathrm{Rb}$ detection and decoupling},}\ }\href {\doibase
  10.1103/PhysRevLett.120.033401} {\bibfield  {journal} {\bibinfo  {journal}
  {Phys. Rev. Lett.}\ }\textbf {\bibinfo {volume} {120}},\ \bibinfo {pages}
  {033401} (\bibinfo {year} {2018})}\BibitemShut {NoStop}%
\bibitem [{\citenamefont {Thrasher}\ \emph {et~al.}(2019)\citenamefont
  {Thrasher}, \citenamefont {Sorensen}, \citenamefont {Weber}, \citenamefont
  {Bulatowicz}, \citenamefont {Korver}, \citenamefont {Larsen},\ and\
  \citenamefont {Walker}}]{NMRgyro2019}%
  \BibitemOpen
  \bibfield  {author} {\bibinfo {author} {\bibfnamefont {D.~A.}\ \bibnamefont
  {Thrasher}}, \bibinfo {author} {\bibfnamefont {S.~S.}\ \bibnamefont
  {Sorensen}}, \bibinfo {author} {\bibfnamefont {J.}~\bibnamefont {Weber}},
  \bibinfo {author} {\bibfnamefont {M.}~\bibnamefont {Bulatowicz}}, \bibinfo
  {author} {\bibfnamefont {A.}~\bibnamefont {Korver}}, \bibinfo {author}
  {\bibfnamefont {M.}~\bibnamefont {Larsen}}, \ and\ \bibinfo {author}
  {\bibfnamefont {T.~G.}\ \bibnamefont {Walker}},\ }\bibfield  {title}
  {\enquote {\bibinfo {title} {Continuous comagnetometry using transversely
  polarized {Xe} isotopes},}\ }\href {\doibase 10.1103/PhysRevA.100.061403}
  {\bibfield  {journal} {\bibinfo  {journal} {Phys. Rev. A}\ }\textbf {\bibinfo
  {volume} {100}},\ \bibinfo {pages} {061403} (\bibinfo {year}
  {2019})}\BibitemShut {NoStop}%
\bibitem [{\citenamefont {Sorensen}\ \emph {et~al.}(2020)\citenamefont
  {Sorensen}, \citenamefont {Thrasher},\ and\ \citenamefont
  {Walker}}]{SORENSEN2020}%
  \BibitemOpen
  \bibfield  {author} {\bibinfo {author} {\bibfnamefont {Susan~S.}\
  \bibnamefont {Sorensen}}, \bibinfo {author} {\bibfnamefont {Daniel~A.}\
  \bibnamefont {Thrasher}}, \ and\ \bibinfo {author} {\bibfnamefont {Thad~G.}\
  \bibnamefont {Walker}},\ }\bibfield  {title} {\enquote {\bibinfo {title} {A
  synchronous spin-exchange optically pumped {NMR}-gyroscope},}\ }\href
  {\doibase 10.3390/app10207099} {\bibfield  {journal} {\bibinfo  {journal}
  {Applied Sciences}\ }\textbf {\bibinfo {volume} {10}} (\bibinfo {year}
  {2020}),\ 10.3390/app10207099}\BibitemShut {NoStop}%
\bibitem [{\citenamefont {Kitching}(2018)}]{KIT2018}%
  \BibitemOpen
  \bibfield  {author} {\bibinfo {author} {\bibfnamefont {John}\ \bibnamefont
  {Kitching}},\ }\bibfield  {title} {\enquote {\bibinfo {title} {Chip-scale
  atomic devices},}\ }\href {\doibase 10.1063/1.5026238} {\bibfield  {journal}
  {\bibinfo  {journal} {Applied Physics Reviews}\ }\textbf {\bibinfo {volume}
  {5}},\ \bibinfo {pages} {031302} (\bibinfo {year} {2018})}\BibitemShut
  {NoStop}%
\bibitem [{\citenamefont {El-Sheimy}\ and\ \citenamefont
  {Youssef}(2020)}]{SHE2020}%
  \BibitemOpen
  \bibfield  {author} {\bibinfo {author} {\bibfnamefont {Naser}\ \bibnamefont
  {El-Sheimy}}\ and\ \bibinfo {author} {\bibfnamefont {Ahmed}\ \bibnamefont
  {Youssef}},\ }\bibfield  {title} {\enquote {\bibinfo {title} {Inertial
  sensors technologies for navigation applications: state of the art and future
  trends},}\ }\href {\doibase 10.1186/s43020-019-0001-5} {\bibfield  {journal}
  {\bibinfo  {journal} {Satellite Navigation}\ }\textbf {\bibinfo {volume}
  {1}},\ \bibinfo {pages} {2} (\bibinfo {year} {2020})}\BibitemShut {NoStop}%
\bibitem [{\citenamefont {Ledbetter}\ \emph {et~al.}(2012)\citenamefont
  {Ledbetter}, \citenamefont {Jensen}, \citenamefont {Fischer}, \citenamefont
  {Jarmola},\ and\ \citenamefont {Budker}}]{LED2012}%
  \BibitemOpen
  \bibfield  {author} {\bibinfo {author} {\bibfnamefont {M.~P.}\ \bibnamefont
  {Ledbetter}}, \bibinfo {author} {\bibfnamefont {K.}~\bibnamefont {Jensen}},
  \bibinfo {author} {\bibfnamefont {R.}~\bibnamefont {Fischer}}, \bibinfo
  {author} {\bibfnamefont {A.}~\bibnamefont {Jarmola}}, \ and\ \bibinfo
  {author} {\bibfnamefont {D.}~\bibnamefont {Budker}},\ }\bibfield  {title}
  {\enquote {\bibinfo {title} {Gyroscopes based on nitrogen-vacancy centers in
  diamond},}\ }\href {\doibase 10.1103/PhysRevA.86.052116} {\bibfield
  {journal} {\bibinfo  {journal} {Phys. Rev. A}\ }\textbf {\bibinfo {volume}
  {86}},\ \bibinfo {pages} {052116} (\bibinfo {year} {2012})}\BibitemShut
  {NoStop}%
\bibitem [{\citenamefont {Ajoy}\ and\ \citenamefont
  {Cappellaro}(2012)}]{AJO2012}%
  \BibitemOpen
  \bibfield  {author} {\bibinfo {author} {\bibfnamefont {Ashok}\ \bibnamefont
  {Ajoy}}\ and\ \bibinfo {author} {\bibfnamefont {Paola}\ \bibnamefont
  {Cappellaro}},\ }\bibfield  {title} {\enquote {\bibinfo {title} {Stable
  three-axis nuclear-spin gyroscope in diamond},}\ }\href {\doibase
  10.1103/PhysRevA.86.062104} {\bibfield  {journal} {\bibinfo  {journal} {Phys.
  Rev. A}\ }\textbf {\bibinfo {volume} {86}},\ \bibinfo {pages} {062104}
  (\bibinfo {year} {2012})}\BibitemShut {NoStop}%
\bibitem [{\citenamefont {Hodges}\ \emph {et~al.}(2013)\citenamefont {Hodges},
  \citenamefont {Yao}, \citenamefont {Maclaurin}, \citenamefont {Rastogi},
  \citenamefont {Lukin},\ and\ \citenamefont {Englund}}]{HOD2013}%
  \BibitemOpen
  \bibfield  {author} {\bibinfo {author} {\bibfnamefont {J.~S.}\ \bibnamefont
  {Hodges}}, \bibinfo {author} {\bibfnamefont {N.~Y.}\ \bibnamefont {Yao}},
  \bibinfo {author} {\bibfnamefont {D.}~\bibnamefont {Maclaurin}}, \bibinfo
  {author} {\bibfnamefont {C.}~\bibnamefont {Rastogi}}, \bibinfo {author}
  {\bibfnamefont {M.~D.}\ \bibnamefont {Lukin}}, \ and\ \bibinfo {author}
  {\bibfnamefont {D.}~\bibnamefont {Englund}},\ }\bibfield  {title} {\enquote
  {\bibinfo {title} {Timekeeping with electron spin states in diamond},}\
  }\href {\doibase 10.1103/PhysRevA.87.032118} {\bibfield  {journal} {\bibinfo
  {journal} {Phys. Rev. A}\ }\textbf {\bibinfo {volume} {87}},\ \bibinfo
  {pages} {032118} (\bibinfo {year} {2013})}\BibitemShut {NoStop}%
\bibitem [{\citenamefont {Fang}\ \emph {et~al.}(2013)\citenamefont {Fang},
  \citenamefont {Acosta}, \citenamefont {Santori}, \citenamefont {Huang},
  \citenamefont {Itoh}, \citenamefont {Watanabe}, \citenamefont {Shikata},\
  and\ \citenamefont {Beausoleil}}]{FAN2013}%
  \BibitemOpen
  \bibfield  {author} {\bibinfo {author} {\bibfnamefont {Kejie}\ \bibnamefont
  {Fang}}, \bibinfo {author} {\bibfnamefont {Victor~M.}\ \bibnamefont
  {Acosta}}, \bibinfo {author} {\bibfnamefont {Charles}\ \bibnamefont
  {Santori}}, \bibinfo {author} {\bibfnamefont {Zhihong}\ \bibnamefont
  {Huang}}, \bibinfo {author} {\bibfnamefont {Kohei~M.}\ \bibnamefont {Itoh}},
  \bibinfo {author} {\bibfnamefont {Hideyuki}\ \bibnamefont {Watanabe}},
  \bibinfo {author} {\bibfnamefont {Shinichi}\ \bibnamefont {Shikata}}, \ and\
  \bibinfo {author} {\bibfnamefont {Raymond~G.}\ \bibnamefont {Beausoleil}},\
  }\bibfield  {title} {\enquote {\bibinfo {title} {High-sensitivity
  magnetometry based on quantum beats in diamond nitrogen-vacancy centers},}\
  }\href {\doibase 10.1103/PhysRevLett.110.130802} {\bibfield  {journal}
  {\bibinfo  {journal} {Phys. Rev. Lett.}\ }\textbf {\bibinfo {volume} {110}},\
  \bibinfo {pages} {130802} (\bibinfo {year} {2013})}\BibitemShut {NoStop}%
\bibitem [{\citenamefont {Degen}\ \emph {et~al.}(2017)\citenamefont {Degen},
  \citenamefont {Reinhard},\ and\ \citenamefont {Cappellaro}}]{DEG2017}%
  \BibitemOpen
  \bibfield  {author} {\bibinfo {author} {\bibfnamefont {C.~L.}\ \bibnamefont
  {Degen}}, \bibinfo {author} {\bibfnamefont {F.}~\bibnamefont {Reinhard}}, \
  and\ \bibinfo {author} {\bibfnamefont {P.}~\bibnamefont {Cappellaro}},\
  }\bibfield  {title} {\enquote {\bibinfo {title} {Quantum sensing},}\ }\href
  {\doibase 10.1103/RevModPhys.89.035002} {\bibfield  {journal} {\bibinfo
  {journal} {Rev. Mod. Phys.}\ }\textbf {\bibinfo {volume} {89}},\ \bibinfo
  {pages} {035002} (\bibinfo {year} {2017})}\BibitemShut {NoStop}%
\bibitem [{\citenamefont {Fu}\ \emph {et~al.}(2020)\citenamefont {Fu},
  \citenamefont {Iwata}, \citenamefont {Wickenbrock},\ and\ \citenamefont
  {Budker}}]{Fu2020_Challenging}%
  \BibitemOpen
  \bibfield  {author} {\bibinfo {author} {\bibfnamefont {Kai-Mei~C.}\
  \bibnamefont {Fu}}, \bibinfo {author} {\bibfnamefont {Geoffrey~Z.}\
  \bibnamefont {Iwata}}, \bibinfo {author} {\bibfnamefont {Arne}\ \bibnamefont
  {Wickenbrock}}, \ and\ \bibinfo {author} {\bibfnamefont {Dmitry}\
  \bibnamefont {Budker}},\ }\bibfield  {title} {\enquote {\bibinfo {title}
  {Sensitive magnetometry in challenging environments},}\ }\href {\doibase
  10.1116/5.0025186} {\bibfield  {journal} {\bibinfo  {journal} {AVS Quantum
  Science}\ }\textbf {\bibinfo {volume} {2}},\ \bibinfo {pages} {044702}
  (\bibinfo {year} {2020})}\BibitemShut {NoStop}%
\bibitem [{\citenamefont {Jarmola}\ \emph {et~al.}(2020)\citenamefont
  {Jarmola}, \citenamefont {Fescenko}, \citenamefont {Acosta}, \citenamefont
  {Doherty}, \citenamefont {Fatemi}, \citenamefont {Ivanov}, \citenamefont
  {Budker},\ and\ \citenamefont {Malinovsky}}]{Jarmola2020Robust}%
  \BibitemOpen
  \bibfield  {author} {\bibinfo {author} {\bibfnamefont {A.}~\bibnamefont
  {Jarmola}}, \bibinfo {author} {\bibfnamefont {I.}~\bibnamefont {Fescenko}},
  \bibinfo {author} {\bibfnamefont {V.~M.}\ \bibnamefont {Acosta}}, \bibinfo
  {author} {\bibfnamefont {M.~W.}\ \bibnamefont {Doherty}}, \bibinfo {author}
  {\bibfnamefont {F.~K.}\ \bibnamefont {Fatemi}}, \bibinfo {author}
  {\bibfnamefont {T.}~\bibnamefont {Ivanov}}, \bibinfo {author} {\bibfnamefont
  {D.}~\bibnamefont {Budker}}, \ and\ \bibinfo {author} {\bibfnamefont {V.~S.}\
  \bibnamefont {Malinovsky}},\ }\bibfield  {title} {\enquote {\bibinfo {title}
  {Robust optical readout and characterization of nuclear spin transitions in
  nitrogen-vacancy ensembles in diamond},}\ }\href {\doibase
  10.1103/PhysRevResearch.2.023094} {\bibfield  {journal} {\bibinfo  {journal}
  {Phys. Rev. Research}\ }\textbf {\bibinfo {volume} {2}},\ \bibinfo {pages}
  {023094} (\bibinfo {year} {2020})}\BibitemShut {NoStop}%
\bibitem [{\citenamefont {Jaskula}\ \emph {et~al.}(2019)\citenamefont
  {Jaskula}, \citenamefont {Saha}, \citenamefont {Ajoy}, \citenamefont
  {Twitchen}, \citenamefont {Markham},\ and\ \citenamefont
  {Cappellaro}}]{JAS2019}%
  \BibitemOpen
  \bibfield  {author} {\bibinfo {author} {\bibfnamefont {J.-C.}\ \bibnamefont
  {Jaskula}}, \bibinfo {author} {\bibfnamefont {K.}~\bibnamefont {Saha}},
  \bibinfo {author} {\bibfnamefont {A.}~\bibnamefont {Ajoy}}, \bibinfo {author}
  {\bibfnamefont {D.J.}\ \bibnamefont {Twitchen}}, \bibinfo {author}
  {\bibfnamefont {M.}~\bibnamefont {Markham}}, \ and\ \bibinfo {author}
  {\bibfnamefont {P.}~\bibnamefont {Cappellaro}},\ }\bibfield  {title}
  {\enquote {\bibinfo {title} {Cross-sensor feedback stabilization of an
  emulated quantum spin gyroscope},}\ }\href {\doibase
  10.1103/PhysRevApplied.11.054010} {\bibfield  {journal} {\bibinfo  {journal}
  {Phys. Rev. Applied}\ }\textbf {\bibinfo {volume} {11}},\ \bibinfo {pages}
  {054010} (\bibinfo {year} {2019})}\BibitemShut {NoStop}%
\bibitem [{\citenamefont {Soshenko}\ \emph {et~al.}(2021)\citenamefont
  {Soshenko}, \citenamefont {Bolshedvorskii}, \citenamefont {Rubinas},
  \citenamefont {Sorokin}, \citenamefont {Smolyaninov}, \citenamefont
  {Vorobyov},\ and\ \citenamefont {Akimov}}]{SOS2020}%
  \BibitemOpen
  \bibfield  {author} {\bibinfo {author} {\bibfnamefont {Vladimir~V.}\
  \bibnamefont {Soshenko}}, \bibinfo {author} {\bibfnamefont {Stepan~V.}\
  \bibnamefont {Bolshedvorskii}}, \bibinfo {author} {\bibfnamefont {Olga}\
  \bibnamefont {Rubinas}}, \bibinfo {author} {\bibfnamefont {Vadim~N.}\
  \bibnamefont {Sorokin}}, \bibinfo {author} {\bibfnamefont {Andrey~N.}\
  \bibnamefont {Smolyaninov}}, \bibinfo {author} {\bibfnamefont {Vadim~V.}\
  \bibnamefont {Vorobyov}}, \ and\ \bibinfo {author} {\bibfnamefont
  {Alexey~V.}\ \bibnamefont {Akimov}},\ }\bibfield  {title} {\enquote {\bibinfo
  {title} {Nuclear spin gyroscope based on the nitrogen vacancy center in
  diamond},}\ }\href {\doibase 10.1103/PhysRevLett.126.197702} {\bibfield
  {journal} {\bibinfo  {journal} {Phys. Rev. Lett.}\ }\textbf {\bibinfo
  {volume} {126}},\ \bibinfo {pages} {197702} (\bibinfo {year}
  {2021})}\BibitemShut {NoStop}%
\bibitem [{\citenamefont {Kraus}\ \emph {et~al.}(2014)\citenamefont {Kraus},
  \citenamefont {Soltamov}, \citenamefont {Fuchs}, \citenamefont {Simin},
  \citenamefont {Sperlich}, \citenamefont {Baranov}, \citenamefont {Astakhov},\
  and\ \citenamefont {Dyakonov}}]{kraus2014}%
  \BibitemOpen
  \bibfield  {author} {\bibinfo {author} {\bibfnamefont {H.}~\bibnamefont
  {Kraus}}, \bibinfo {author} {\bibfnamefont {V.~A.}\ \bibnamefont {Soltamov}},
  \bibinfo {author} {\bibfnamefont {F.}~\bibnamefont {Fuchs}}, \bibinfo
  {author} {\bibfnamefont {D.}~\bibnamefont {Simin}}, \bibinfo {author}
  {\bibfnamefont {A.}~\bibnamefont {Sperlich}}, \bibinfo {author}
  {\bibfnamefont {P.~G.}\ \bibnamefont {Baranov}}, \bibinfo {author}
  {\bibfnamefont {G.~V.}\ \bibnamefont {Astakhov}}, \ and\ \bibinfo {author}
  {\bibfnamefont {V.}~\bibnamefont {Dyakonov}},\ }\bibfield  {title} {\enquote
  {\bibinfo {title} {Magnetic field and temperature sensing with atomic-scale
  spin defects in silicon carbide},}\ }\href {\doibase 10.1038/srep05303}
  {\bibfield  {journal} {\bibinfo  {journal} {Scientific Reports}\ }\textbf
  {\bibinfo {volume} {4}},\ \bibinfo {pages} {5303} (\bibinfo {year}
  {2014})}\BibitemShut {NoStop}%
\bibitem [{\citenamefont {Bulatowicz}\ \emph {et~al.}(2013)\citenamefont
  {Bulatowicz}, \citenamefont {Griffith}, \citenamefont {Larsen}, \citenamefont
  {Mirijanian}, \citenamefont {Fu}, \citenamefont {Smith}, \citenamefont
  {Snow}, \citenamefont {Yan},\ and\ \citenamefont {Walker}}]{BUL2013}%
  \BibitemOpen
  \bibfield  {author} {\bibinfo {author} {\bibfnamefont {M.}~\bibnamefont
  {Bulatowicz}}, \bibinfo {author} {\bibfnamefont {R.}~\bibnamefont
  {Griffith}}, \bibinfo {author} {\bibfnamefont {M.}~\bibnamefont {Larsen}},
  \bibinfo {author} {\bibfnamefont {J.}~\bibnamefont {Mirijanian}}, \bibinfo
  {author} {\bibfnamefont {C.~B.}\ \bibnamefont {Fu}}, \bibinfo {author}
  {\bibfnamefont {E.}~\bibnamefont {Smith}}, \bibinfo {author} {\bibfnamefont
  {W.~M.}\ \bibnamefont {Snow}}, \bibinfo {author} {\bibfnamefont
  {H.}~\bibnamefont {Yan}}, \ and\ \bibinfo {author} {\bibfnamefont {T.~G.}\
  \bibnamefont {Walker}},\ }\bibfield  {title} {\enquote {\bibinfo {title}
  {Laboratory search for a long-range $t$-odd, $p$-odd interaction from
  axionlike particles using dual-species nuclear magnetic resonance with
  polarized $^{129}\mathrm{Xe}$ and $^{131}\mathrm{Xe}$ gas},}\ }\href
  {\doibase 10.1103/PhysRevLett.111.102001} {\bibfield  {journal} {\bibinfo
  {journal} {Phys. Rev. Lett.}\ }\textbf {\bibinfo {volume} {111}},\ \bibinfo
  {pages} {102001} (\bibinfo {year} {2013})}\BibitemShut {NoStop}%
\bibitem [{\citenamefont {Rong}\ \emph {et~al.}(2018)\citenamefont {Rong},
  \citenamefont {Wang}, \citenamefont {Geng}, \citenamefont {Qin},
  \citenamefont {Guo}, \citenamefont {Jiao}, \citenamefont {Xie}, \citenamefont
  {Wang}, \citenamefont {Huang}, \citenamefont {Shi}, \citenamefont {Cai},
  \citenamefont {Zou},\ and\ \citenamefont {Du}}]{Rong2018}%
  \BibitemOpen
  \bibfield  {author} {\bibinfo {author} {\bibfnamefont {Xing}\ \bibnamefont
  {Rong}}, \bibinfo {author} {\bibfnamefont {Mengqi}\ \bibnamefont {Wang}},
  \bibinfo {author} {\bibfnamefont {Jianpei}\ \bibnamefont {Geng}}, \bibinfo
  {author} {\bibfnamefont {Xi}~\bibnamefont {Qin}}, \bibinfo {author}
  {\bibfnamefont {Maosen}\ \bibnamefont {Guo}}, \bibinfo {author}
  {\bibfnamefont {Man}\ \bibnamefont {Jiao}}, \bibinfo {author} {\bibfnamefont
  {Yijin}\ \bibnamefont {Xie}}, \bibinfo {author} {\bibfnamefont {Pengfei}\
  \bibnamefont {Wang}}, \bibinfo {author} {\bibfnamefont {Pu}~\bibnamefont
  {Huang}}, \bibinfo {author} {\bibfnamefont {Fazhan}\ \bibnamefont {Shi}},
  \bibinfo {author} {\bibfnamefont {Yi-Fu}\ \bibnamefont {Cai}}, \bibinfo
  {author} {\bibfnamefont {Chongwen}\ \bibnamefont {Zou}}, \ and\ \bibinfo
  {author} {\bibfnamefont {Jiangfeng}\ \bibnamefont {Du}},\ }\bibfield  {title}
  {\enquote {\bibinfo {title} {Searching for an exotic spin-dependent
  interaction with a single electron-spin quantum sensor},}\ }\href {\doibase
  10.1038/s41467-018-03152-9} {\bibfield  {journal} {\bibinfo  {journal}
  {Nature Communications}\ }\textbf {\bibinfo {volume} {9}},\ \bibinfo {pages}
  {739} (\bibinfo {year} {2018})}\BibitemShut {NoStop}%
\bibitem [{\citenamefont {Ding}\ \emph {et~al.}(2020)\citenamefont {Ding},
  \citenamefont {Wang}, \citenamefont {Zhou}, \citenamefont {Liu},
  \citenamefont {Sun}, \citenamefont {Adeyeye}, \citenamefont {Fu},
  \citenamefont {Ren}, \citenamefont {Li}, \citenamefont {Luo}, \citenamefont
  {Lan}, \citenamefont {Yang},\ and\ \citenamefont {Luo}}]{DING2020}%
  \BibitemOpen
  \bibfield  {author} {\bibinfo {author} {\bibfnamefont {Jihua}\ \bibnamefont
  {Ding}}, \bibinfo {author} {\bibfnamefont {Jianbo}\ \bibnamefont {Wang}},
  \bibinfo {author} {\bibfnamefont {Xue}\ \bibnamefont {Zhou}}, \bibinfo
  {author} {\bibfnamefont {Yu}~\bibnamefont {Liu}}, \bibinfo {author}
  {\bibfnamefont {Ke}~\bibnamefont {Sun}}, \bibinfo {author} {\bibfnamefont
  {Adekunle~Olusola}\ \bibnamefont {Adeyeye}}, \bibinfo {author} {\bibfnamefont
  {Huixing}\ \bibnamefont {Fu}}, \bibinfo {author} {\bibfnamefont {Xiaofang}\
  \bibnamefont {Ren}}, \bibinfo {author} {\bibfnamefont {Sumin}\ \bibnamefont
  {Li}}, \bibinfo {author} {\bibfnamefont {Pengshun}\ \bibnamefont {Luo}},
  \bibinfo {author} {\bibfnamefont {Zhongwen}\ \bibnamefont {Lan}}, \bibinfo
  {author} {\bibfnamefont {Shanqing}\ \bibnamefont {Yang}}, \ and\ \bibinfo
  {author} {\bibfnamefont {Jun}\ \bibnamefont {Luo}},\ }\bibfield  {title}
  {\enquote {\bibinfo {title} {Constraints on the velocity and spin dependent
  exotic interaction at the micrometer range},}\ }\href {\doibase
  10.1103/PhysRevLett.124.161801} {\bibfield  {journal} {\bibinfo  {journal}
  {Phys. Rev. Lett.}\ }\textbf {\bibinfo {volume} {124}},\ \bibinfo {pages}
  {161801} (\bibinfo {year} {2020})}\BibitemShut {NoStop}%
\bibitem [{\citenamefont {Jacques}\ \emph {et~al.}(2009)\citenamefont
  {Jacques}, \citenamefont {Neumann}, \citenamefont {Beck}, \citenamefont
  {Markham}, \citenamefont {Twitchen}, \citenamefont {Meijer}, \citenamefont
  {Kaiser}, \citenamefont {Balasubramanian}, \citenamefont {Jelezko},\ and\
  \citenamefont {Wrachtrup}}]{JAC2009}%
  \BibitemOpen
  \bibfield  {author} {\bibinfo {author} {\bibfnamefont {V.}~\bibnamefont
  {Jacques}}, \bibinfo {author} {\bibfnamefont {P.}~\bibnamefont {Neumann}},
  \bibinfo {author} {\bibfnamefont {J.}~\bibnamefont {Beck}}, \bibinfo {author}
  {\bibfnamefont {M.}~\bibnamefont {Markham}}, \bibinfo {author} {\bibfnamefont
  {D.}~\bibnamefont {Twitchen}}, \bibinfo {author} {\bibfnamefont
  {J.}~\bibnamefont {Meijer}}, \bibinfo {author} {\bibfnamefont
  {F.}~\bibnamefont {Kaiser}}, \bibinfo {author} {\bibfnamefont
  {G.}~\bibnamefont {Balasubramanian}}, \bibinfo {author} {\bibfnamefont
  {F.}~\bibnamefont {Jelezko}}, \ and\ \bibinfo {author} {\bibfnamefont
  {J.}~\bibnamefont {Wrachtrup}},\ }\bibfield  {title} {\enquote {\bibinfo
  {title} {Dynamic polarization of single nuclear spins by optical pumping of
  nitrogen-vacancy color centers in diamond at room temperature},}\ }\href
  {\doibase 10.1103/PhysRevLett.102.057403} {\bibfield  {journal} {\bibinfo
  {journal} {Phys. Rev. Lett.}\ }\textbf {\bibinfo {volume} {102}},\ \bibinfo
  {pages} {057403} (\bibinfo {year} {2009})}\BibitemShut {NoStop}%
\bibitem [{\citenamefont {Smeltzer}\ \emph {et~al.}(2009)\citenamefont
  {Smeltzer}, \citenamefont {McIntyre},\ and\ \citenamefont
  {Childress}}]{SME2009}%
  \BibitemOpen
  \bibfield  {author} {\bibinfo {author} {\bibfnamefont {Benjamin}\
  \bibnamefont {Smeltzer}}, \bibinfo {author} {\bibfnamefont {Jean}\
  \bibnamefont {McIntyre}}, \ and\ \bibinfo {author} {\bibfnamefont {Lilian}\
  \bibnamefont {Childress}},\ }\bibfield  {title} {\enquote {\bibinfo {title}
  {Robust control of individual nuclear spins in diamond},}\ }\href {\doibase
  10.1103/PhysRevA.80.050302} {\bibfield  {journal} {\bibinfo  {journal} {Phys.
  Rev. A}\ }\textbf {\bibinfo {volume} {80}},\ \bibinfo {pages} {050302}
  (\bibinfo {year} {2009})}\BibitemShut {NoStop}%
\bibitem [{\citenamefont {Steiner}\ \emph {et~al.}(2010)\citenamefont
  {Steiner}, \citenamefont {Neumann}, \citenamefont {Beck}, \citenamefont
  {Jelezko},\ and\ \citenamefont {Wrachtrup}}]{STE2010PRB}%
  \BibitemOpen
  \bibfield  {author} {\bibinfo {author} {\bibfnamefont {M.}~\bibnamefont
  {Steiner}}, \bibinfo {author} {\bibfnamefont {P.}~\bibnamefont {Neumann}},
  \bibinfo {author} {\bibfnamefont {J.}~\bibnamefont {Beck}}, \bibinfo {author}
  {\bibfnamefont {F.}~\bibnamefont {Jelezko}}, \ and\ \bibinfo {author}
  {\bibfnamefont {J.}~\bibnamefont {Wrachtrup}},\ }\bibfield  {title} {\enquote
  {\bibinfo {title} {Universal enhancement of the optical readout fidelity of
  single electron spins at nitrogen-vacancy centers in diamond},}\ }\href
  {\doibase 10.1103/PhysRevB.81.035205} {\bibfield  {journal} {\bibinfo
  {journal} {Phys. Rev. B}\ }\textbf {\bibinfo {volume} {81}},\ \bibinfo
  {pages} {035205} (\bibinfo {year} {2010})}\BibitemShut {NoStop}%
\bibitem [{\citenamefont {Fischer}\ \emph {et~al.}(2013)\citenamefont
  {Fischer}, \citenamefont {Jarmola}, \citenamefont {Kehayias},\ and\
  \citenamefont {Budker}}]{FIS2013PRB}%
  \BibitemOpen
  \bibfield  {author} {\bibinfo {author} {\bibfnamefont {Ran}\ \bibnamefont
  {Fischer}}, \bibinfo {author} {\bibfnamefont {Andrey}\ \bibnamefont
  {Jarmola}}, \bibinfo {author} {\bibfnamefont {Pauli}\ \bibnamefont
  {Kehayias}}, \ and\ \bibinfo {author} {\bibfnamefont {Dmitry}\ \bibnamefont
  {Budker}},\ }\bibfield  {title} {\enquote {\bibinfo {title} {Optical
  polarization of nuclear ensembles in diamond},}\ }\href {\doibase
  10.1103/PhysRevB.87.125207} {\bibfield  {journal} {\bibinfo  {journal} {Phys.
  Rev. B}\ }\textbf {\bibinfo {volume} {87}},\ \bibinfo {pages} {125207}
  (\bibinfo {year} {2013})}\BibitemShut {NoStop}%
\bibitem [{\citenamefont {Hart}\ \emph {et~al.}(2021)\citenamefont {Hart},
  \citenamefont {Schloss}, \citenamefont {Turner}, \citenamefont {Scheidegger},
  \citenamefont {Bauch},\ and\ \citenamefont {Walsworth}}]{Hart2020}%
  \BibitemOpen
  \bibfield  {author} {\bibinfo {author} {\bibfnamefont {Connor~A.}\
  \bibnamefont {Hart}}, \bibinfo {author} {\bibfnamefont {Jennifer~M.}\
  \bibnamefont {Schloss}}, \bibinfo {author} {\bibfnamefont {Matthew~J.}\
  \bibnamefont {Turner}}, \bibinfo {author} {\bibfnamefont {Patrick~J.}\
  \bibnamefont {Scheidegger}}, \bibinfo {author} {\bibfnamefont {Erik}\
  \bibnamefont {Bauch}}, \ and\ \bibinfo {author} {\bibfnamefont {Ronald~L.}\
  \bibnamefont {Walsworth}},\ }\bibfield  {title} {\enquote {\bibinfo {title}
  {{N}-{V}--diamond magnetic microscopy using a double quantum 4-ramsey
  protocol},}\ }\href {\doibase 10.1103/PhysRevApplied.15.044020} {\bibfield
  {journal} {\bibinfo  {journal} {Phys. Rev. Applied}\ }\textbf {\bibinfo
  {volume} {15}},\ \bibinfo {pages} {044020} (\bibinfo {year}
  {2021})}\BibitemShut {NoStop}%
\bibitem [{\citenamefont {Le~Sage}\ \emph {et~al.}(2012)\citenamefont
  {Le~Sage}, \citenamefont {Pham}, \citenamefont {Bar-Gill}, \citenamefont
  {Belthangady}, \citenamefont {Lukin}, \citenamefont {Yacoby},\ and\
  \citenamefont {Walsworth}}]{SAG2012}%
  \BibitemOpen
  \bibfield  {author} {\bibinfo {author} {\bibfnamefont {D.}~\bibnamefont
  {Le~Sage}}, \bibinfo {author} {\bibfnamefont {L.~M.}\ \bibnamefont {Pham}},
  \bibinfo {author} {\bibfnamefont {N.}~\bibnamefont {Bar-Gill}}, \bibinfo
  {author} {\bibfnamefont {C.}~\bibnamefont {Belthangady}}, \bibinfo {author}
  {\bibfnamefont {M.~D.}\ \bibnamefont {Lukin}}, \bibinfo {author}
  {\bibfnamefont {A.}~\bibnamefont {Yacoby}}, \ and\ \bibinfo {author}
  {\bibfnamefont {R.~L.}\ \bibnamefont {Walsworth}},\ }\bibfield  {title}
  {\enquote {\bibinfo {title} {Efficient photon detection from color centers in
  a diamond optical waveguide},}\ }\href {\doibase 10.1103/PhysRevB.85.121202}
  {\bibfield  {journal} {\bibinfo  {journal} {Phys. Rev. B}\ }\textbf {\bibinfo
  {volume} {85}},\ \bibinfo {pages} {121202} (\bibinfo {year}
  {2012})}\BibitemShut {NoStop}%
\bibitem [{\citenamefont {Wolf}\ \emph {et~al.}(2015)\citenamefont {Wolf},
  \citenamefont {Neumann}, \citenamefont {Nakamura}, \citenamefont {Sumiya},
  \citenamefont {Ohshima}, \citenamefont {Isoya},\ and\ \citenamefont
  {Wrachtrup}}]{WOL2015}%
  \BibitemOpen
  \bibfield  {author} {\bibinfo {author} {\bibfnamefont {Thomas}\ \bibnamefont
  {Wolf}}, \bibinfo {author} {\bibfnamefont {Philipp}\ \bibnamefont {Neumann}},
  \bibinfo {author} {\bibfnamefont {Kazuo}\ \bibnamefont {Nakamura}}, \bibinfo
  {author} {\bibfnamefont {Hitoshi}\ \bibnamefont {Sumiya}}, \bibinfo {author}
  {\bibfnamefont {Takeshi}\ \bibnamefont {Ohshima}}, \bibinfo {author}
  {\bibfnamefont {Junichi}\ \bibnamefont {Isoya}}, \ and\ \bibinfo {author}
  {\bibfnamefont {J\"org}\ \bibnamefont {Wrachtrup}},\ }\bibfield  {title}
  {\enquote {\bibinfo {title} {Subpicotesla diamond magnetometry},}\ }\href
  {\doibase 10.1103/PhysRevX.5.041001} {\bibfield  {journal} {\bibinfo
  {journal} {Phys. Rev. X}\ }\textbf {\bibinfo {volume} {5}},\ \bibinfo {pages}
  {041001} (\bibinfo {year} {2015})}\BibitemShut {NoStop}%
\bibitem [{\citenamefont {de~Lange}\ \emph {et~al.}(2012)\citenamefont
  {de~Lange}, \citenamefont {van~der Sar}, \citenamefont {Blok}, \citenamefont
  {Wang}, \citenamefont {Dobrovitski},\ and\ \citenamefont {Hanson}}]{DEL2012}%
  \BibitemOpen
  \bibfield  {author} {\bibinfo {author} {\bibfnamefont {Gijs}\ \bibnamefont
  {de~Lange}}, \bibinfo {author} {\bibfnamefont {Toeno}\ \bibnamefont {van~der
  Sar}}, \bibinfo {author} {\bibfnamefont {Machiel}\ \bibnamefont {Blok}},
  \bibinfo {author} {\bibfnamefont {Zhi-Hui}\ \bibnamefont {Wang}}, \bibinfo
  {author} {\bibfnamefont {Viatcheslav}\ \bibnamefont {Dobrovitski}}, \ and\
  \bibinfo {author} {\bibfnamefont {Ronald}\ \bibnamefont {Hanson}},\
  }\bibfield  {title} {\enquote {\bibinfo {title} {Controlling the quantum
  dynamics of a mesoscopic spin bath in diamond},}\ }\href {\doibase
  10.1038/srep00382} {\bibfield  {journal} {\bibinfo  {journal} {Scientific
  Reports}\ }\textbf {\bibinfo {volume} {2}},\ \bibinfo {pages} {382} (\bibinfo
  {year} {2012})}\BibitemShut {NoStop}%
\bibitem [{\citenamefont {Bauch}\ \emph {et~al.}(2018)\citenamefont {Bauch},
  \citenamefont {Hart}, \citenamefont {Schloss}, \citenamefont {Turner},
  \citenamefont {Barry}, \citenamefont {Kehayias}, \citenamefont {Singh},\ and\
  \citenamefont {Walsworth}}]{BAU2018}%
  \BibitemOpen
  \bibfield  {author} {\bibinfo {author} {\bibfnamefont {Erik}\ \bibnamefont
  {Bauch}}, \bibinfo {author} {\bibfnamefont {Connor~A.}\ \bibnamefont {Hart}},
  \bibinfo {author} {\bibfnamefont {Jennifer~M.}\ \bibnamefont {Schloss}},
  \bibinfo {author} {\bibfnamefont {Matthew~J.}\ \bibnamefont {Turner}},
  \bibinfo {author} {\bibfnamefont {John~F.}\ \bibnamefont {Barry}}, \bibinfo
  {author} {\bibfnamefont {Pauli}\ \bibnamefont {Kehayias}}, \bibinfo {author}
  {\bibfnamefont {Swati}\ \bibnamefont {Singh}}, \ and\ \bibinfo {author}
  {\bibfnamefont {Ronald~L.}\ \bibnamefont {Walsworth}},\ }\bibfield  {title}
  {\enquote {\bibinfo {title} {Ultralong dephasing times in solid-state spin
  ensembles via quantum control},}\ }\href {\doibase 10.1103/PhysRevX.8.031025}
  {\bibfield  {journal} {\bibinfo  {journal} {Phys. Rev. X}\ }\textbf {\bibinfo
  {volume} {8}},\ \bibinfo {pages} {031025} (\bibinfo {year}
  {2018})}\BibitemShut {NoStop}%
\bibitem [{\citenamefont {Alvarez}\ \emph {et~al.}(2015)\citenamefont
  {Alvarez}, \citenamefont {Bretschneider}, \citenamefont {Fischer},
  \citenamefont {London}, \citenamefont {Kanda}, \citenamefont {Onoda},
  \citenamefont {Isoya}, \citenamefont {Gershoni},\ and\ \citenamefont
  {Frydman}}]{ALV2015}%
  \BibitemOpen
  \bibfield  {author} {\bibinfo {author} {\bibfnamefont {Gonzalo~A.}\
  \bibnamefont {Alvarez}}, \bibinfo {author} {\bibfnamefont {Christian~O.}\
  \bibnamefont {Bretschneider}}, \bibinfo {author} {\bibfnamefont {Ran}\
  \bibnamefont {Fischer}}, \bibinfo {author} {\bibfnamefont {Paz}\ \bibnamefont
  {London}}, \bibinfo {author} {\bibfnamefont {Hisao}\ \bibnamefont {Kanda}},
  \bibinfo {author} {\bibfnamefont {Shinobu}\ \bibnamefont {Onoda}}, \bibinfo
  {author} {\bibfnamefont {Junichi}\ \bibnamefont {Isoya}}, \bibinfo {author}
  {\bibfnamefont {David}\ \bibnamefont {Gershoni}}, \ and\ \bibinfo {author}
  {\bibfnamefont {Lucio}\ \bibnamefont {Frydman}},\ }\bibfield  {title}
  {\enquote {\bibinfo {title} {Local and bulk $^{13}\mathrm{C}$
  hyperpolarization in nitrogen-vacancy-centred diamonds at variable fields and
  orientations},}\ }\href {\doibase doi:10.1038/ncomms9456} {\bibfield
  {journal} {\bibinfo  {journal} {Nature Communications}\ }\textbf {\bibinfo
  {volume} {6}} (\bibinfo {year} {2015}),\ doi:10.1038/ncomms9456}\BibitemShut
  {NoStop}%
\bibitem [{\citenamefont {King}\ \emph {et~al.}(2015)\citenamefont {King},
  \citenamefont {Jeong}, \citenamefont {Vassiliou}, \citenamefont {Shin},
  \citenamefont {Page}, \citenamefont {Avalos}, \citenamefont {Wang},\ and\
  \citenamefont {Pines}}]{KIN2015}%
  \BibitemOpen
  \bibfield  {author} {\bibinfo {author} {\bibfnamefont {Jonathan~P.}\
  \bibnamefont {King}}, \bibinfo {author} {\bibfnamefont {Keunhong}\
  \bibnamefont {Jeong}}, \bibinfo {author} {\bibfnamefont {Christophoros~C.}\
  \bibnamefont {Vassiliou}}, \bibinfo {author} {\bibfnamefont {Chang~S.}\
  \bibnamefont {Shin}}, \bibinfo {author} {\bibfnamefont {Ralph~H.}\
  \bibnamefont {Page}}, \bibinfo {author} {\bibfnamefont {Claudia~E.}\
  \bibnamefont {Avalos}}, \bibinfo {author} {\bibfnamefont {Hai-Jing}\
  \bibnamefont {Wang}}, \ and\ \bibinfo {author} {\bibfnamefont {Alexander}\
  \bibnamefont {Pines}},\ }\bibfield  {title} {\enquote {\bibinfo {title}
  {Room-temperature in situ nuclear spin hyperpolarization from optically
  pumped nitrogen vacancy centres in diamond},}\ }\href {\doibase
  10.1038/ncomms9965} {\bibfield  {journal} {\bibinfo  {journal} {Nature
  Communications}\ }\textbf {\bibinfo {volume} {6}} (\bibinfo {year} {2015}),\
  10.1038/ncomms9965}\BibitemShut {NoStop}%
\bibitem [{\citenamefont {Ajoy}\ \emph {et~al.}(2018)\citenamefont {Ajoy},
  \citenamefont {Liu}, \citenamefont {Nazaryan}, \citenamefont {Lv},
  \citenamefont {Zangara}, \citenamefont {Safvati}, \citenamefont {Wang},
  \citenamefont {Arnold}, \citenamefont {Li}, \citenamefont {Lin},
  \citenamefont {Raghavan}, \citenamefont {Druga}, \citenamefont {Dhomkar},
  \citenamefont {Pagliero}, \citenamefont {Reimer}, \citenamefont {Suter},
  \citenamefont {Meriles},\ and\ \citenamefont {Pines}}]{Ajoy2018}%
  \BibitemOpen
  \bibfield  {author} {\bibinfo {author} {\bibfnamefont {Ashok}\ \bibnamefont
  {Ajoy}}, \bibinfo {author} {\bibfnamefont {Kristina}\ \bibnamefont {Liu}},
  \bibinfo {author} {\bibfnamefont {Raffi}\ \bibnamefont {Nazaryan}}, \bibinfo
  {author} {\bibfnamefont {Xudong}\ \bibnamefont {Lv}}, \bibinfo {author}
  {\bibfnamefont {Pablo~R.}\ \bibnamefont {Zangara}}, \bibinfo {author}
  {\bibfnamefont {Benjamin}\ \bibnamefont {Safvati}}, \bibinfo {author}
  {\bibfnamefont {Guoqing}\ \bibnamefont {Wang}}, \bibinfo {author}
  {\bibfnamefont {Daniel}\ \bibnamefont {Arnold}}, \bibinfo {author}
  {\bibfnamefont {Grace}\ \bibnamefont {Li}}, \bibinfo {author} {\bibfnamefont
  {Arthur}\ \bibnamefont {Lin}}, \bibinfo {author} {\bibfnamefont {Priyanka}\
  \bibnamefont {Raghavan}}, \bibinfo {author} {\bibfnamefont {Emanuel}\
  \bibnamefont {Druga}}, \bibinfo {author} {\bibfnamefont {Siddharth}\
  \bibnamefont {Dhomkar}}, \bibinfo {author} {\bibfnamefont {Daniela}\
  \bibnamefont {Pagliero}}, \bibinfo {author} {\bibfnamefont {Jeffrey~A.}\
  \bibnamefont {Reimer}}, \bibinfo {author} {\bibfnamefont {Dieter}\
  \bibnamefont {Suter}}, \bibinfo {author} {\bibfnamefont {Carlos~A.}\
  \bibnamefont {Meriles}}, \ and\ \bibinfo {author} {\bibfnamefont {Alexander}\
  \bibnamefont {Pines}},\ }\bibfield  {title} {\enquote {\bibinfo {title}
  {Orientation-independent room temperature optical $^{13}\mathrm{C}$
  hyperpolarization in powdered diamond},}\ }\href {\doibase
  10.1126/sciadv.aar5492} {\bibfield  {journal} {\bibinfo  {journal} {Science
  Advances}\ }\textbf {\bibinfo {volume} {4}} (\bibinfo {year} {2018}),\
  10.1126/sciadv.aar5492}\BibitemShut {NoStop}%
\bibitem [{\citenamefont {Pagliero}\ \emph {et~al.}(2018)\citenamefont
  {Pagliero}, \citenamefont {Rao}, \citenamefont {Zangara}, \citenamefont
  {Dhomkar}, \citenamefont {Wong}, \citenamefont {Abril}, \citenamefont
  {Aslam}, \citenamefont {Parker}, \citenamefont {King}, \citenamefont
  {Avalos}, \citenamefont {Ajoy}, \citenamefont {Wrachtrup}, \citenamefont
  {Pines},\ and\ \citenamefont {Meriles}}]{Pagliero2018}%
  \BibitemOpen
  \bibfield  {author} {\bibinfo {author} {\bibfnamefont {Daniela}\ \bibnamefont
  {Pagliero}}, \bibinfo {author} {\bibfnamefont {K.~R.~Koteswara}\ \bibnamefont
  {Rao}}, \bibinfo {author} {\bibfnamefont {Pablo~R.}\ \bibnamefont {Zangara}},
  \bibinfo {author} {\bibfnamefont {Siddharth}\ \bibnamefont {Dhomkar}},
  \bibinfo {author} {\bibfnamefont {Henry~H.}\ \bibnamefont {Wong}}, \bibinfo
  {author} {\bibfnamefont {Andrea}\ \bibnamefont {Abril}}, \bibinfo {author}
  {\bibfnamefont {Nabeel}\ \bibnamefont {Aslam}}, \bibinfo {author}
  {\bibfnamefont {Anna}\ \bibnamefont {Parker}}, \bibinfo {author}
  {\bibfnamefont {Jonathan}\ \bibnamefont {King}}, \bibinfo {author}
  {\bibfnamefont {Claudia~E.}\ \bibnamefont {Avalos}}, \bibinfo {author}
  {\bibfnamefont {Ashok}\ \bibnamefont {Ajoy}}, \bibinfo {author}
  {\bibfnamefont {Joerg}\ \bibnamefont {Wrachtrup}}, \bibinfo {author}
  {\bibfnamefont {Alexander}\ \bibnamefont {Pines}}, \ and\ \bibinfo {author}
  {\bibfnamefont {Carlos~A.}\ \bibnamefont {Meriles}},\ }\bibfield  {title}
  {\enquote {\bibinfo {title} {Multispin-assisted optical pumping of bulk
  $^{13}\mathrm{C}$ nuclear spin polarization in diamond},}\ }\href {\doibase
  10.1103/PhysRevB.97.024422} {\bibfield  {journal} {\bibinfo  {journal} {Phys.
  Rev. B}\ }\textbf {\bibinfo {volume} {97}},\ \bibinfo {pages} {024422}
  (\bibinfo {year} {2018})}\BibitemShut {NoStop}%
\bibitem [{\citenamefont {Scheuer}\ \emph {et~al.}(2017)\citenamefont
  {Scheuer}, \citenamefont {Schwartz}, \citenamefont {M\"uller}, \citenamefont
  {Chen}, \citenamefont {Dhand}, \citenamefont {Plenio}, \citenamefont
  {Naydenov},\ and\ \citenamefont {Jelezko}}]{Sch2017}%
  \BibitemOpen
  \bibfield  {author} {\bibinfo {author} {\bibfnamefont {Jochen}\ \bibnamefont
  {Scheuer}}, \bibinfo {author} {\bibfnamefont {Ilai}\ \bibnamefont
  {Schwartz}}, \bibinfo {author} {\bibfnamefont {Samuel}\ \bibnamefont
  {M\"uller}}, \bibinfo {author} {\bibfnamefont {Qiong}\ \bibnamefont {Chen}},
  \bibinfo {author} {\bibfnamefont {Ish}\ \bibnamefont {Dhand}}, \bibinfo
  {author} {\bibfnamefont {Martin~B.}\ \bibnamefont {Plenio}}, \bibinfo
  {author} {\bibfnamefont {Boris}\ \bibnamefont {Naydenov}}, \ and\ \bibinfo
  {author} {\bibfnamefont {Fedor}\ \bibnamefont {Jelezko}},\ }\bibfield
  {title} {\enquote {\bibinfo {title} {Robust techniques for polarization and
  detection of nuclear spin ensembles},}\ }\href {\doibase
  10.1103/PhysRevB.96.174436} {\bibfield  {journal} {\bibinfo  {journal} {Phys.
  Rev. B}\ }\textbf {\bibinfo {volume} {96}},\ \bibinfo {pages} {174436}
  (\bibinfo {year} {2017})}\BibitemShut {NoStop}%
\bibitem [{\citenamefont {Wood}\ \emph {et~al.}(2017)\citenamefont {Wood},
  \citenamefont {Lilette}, \citenamefont {Fein}, \citenamefont {Perunicic},
  \citenamefont {Hollenberg}, \citenamefont {Scholten},\ and\ \citenamefont
  {Martin}}]{Wood2017}%
  \BibitemOpen
  \bibfield  {author} {\bibinfo {author} {\bibfnamefont {A.~ ~A.}\
  \bibnamefont {Wood}}, \bibinfo {author} {\bibfnamefont {E.}~\bibnamefont
  {Lilette}}, \bibinfo {author} {\bibfnamefont {Y.~Y.}\ \bibnamefont {Fein}},
  \bibinfo {author} {\bibfnamefont {V.~S.}\ \bibnamefont {Perunicic}}, \bibinfo
  {author} {\bibfnamefont {L.~  C.  ~L.}\ \bibnamefont {Hollenberg}},
  \bibinfo {author} {\bibfnamefont {R.~E.}\ \bibnamefont {Scholten}}, \ and\
  \bibinfo {author} {\bibfnamefont {A.~M.}\ \bibnamefont {Martin}},\ }\bibfield
   {title} {\enquote {\bibinfo {title} {Magnetic pseudo-fields in a rotating
  electron--nuclear spin system},}\ }\href {\doibase 10.1038/nphys4221}
  {\bibfield  {journal} {\bibinfo  {journal} {Nature Physics}\ }\textbf
  {\bibinfo {volume} {13}},\ \bibinfo {pages} {1070--1073} (\bibinfo {year}
  {2017})}\BibitemShut {NoStop}%
\end{thebibliography}

\begin{thebibliography}{1}%
\makeatletter
\providecommand \@ifxundefined [1]{%
 \@ifx{#1\undefined}
}%
\providecommand \@ifnum [1]{%
 \ifnum #1\expandafter \@firstoftwo
 \else \expandafter \@secondoftwo
 \fi
}%
\providecommand \@ifx [1]{%
 \ifx #1\expandafter \@firstoftwo
 \else \expandafter \@secondoftwo
 \fi
}%
\providecommand \natexlab [1]{#1}%
\providecommand \enquote  [1]{``#1''}%
\providecommand \bibnamefont  [1]{#1}%
\providecommand \bibfnamefont [1]{#1}%
\providecommand \citenamefont [1]{#1}%
\providecommand \href@noop [0]{\@secondoftwo}%
\providecommand \href [0]{\begingroup \@sanitize@url \@href}%
\providecommand \@href[1]{\@@startlink{#1}\@@href}%
\providecommand \@@href[1]{\endgroup#1\@@endlink}%
\providecommand \@sanitize@url [0]{\catcode `\\12\catcode `\$12\catcode
  `\&12\catcode `\#12\catcode `\^12\catcode `\_12\catcode `\%12\relax}%
\providecommand \@@startlink[1]{}%
\providecommand \@@endlink[0]{}%
\providecommand \url  [0]{\begingroup\@sanitize@url \@url }%
\providecommand \@url [1]{\endgroup\@href {#1}{\urlprefix }}%
\providecommand \urlprefix  [0]{URL }%
\providecommand \Eprint [0]{\href }%
\providecommand \doibase [0]{http://dx.doi.org/}%
\providecommand \selectlanguage [0]{\@gobble}%
\providecommand \bibinfo  [0]{\@secondoftwo}%
\providecommand \bibfield  [0]{\@secondoftwo}%
\providecommand \translation [1]{[#1]}%
\providecommand \BibitemOpen [0]{}%
\providecommand \bibitemStop [0]{}%
\providecommand \bibitemNoStop [0]{.\EOS\space}%
\providecommand \EOS [0]{\spacefactor3000\relax}%
\providecommand \BibitemShut  [1]{\csname bibitem#1\endcsname}%
\let\auto@bib@innerbib\@empty
%</preamble>
\bibitem [{\citenamefont {Jarmola}\ \emph {et~al.}(2020)\citenamefont
  {Jarmola}, \citenamefont {Fescenko}, \citenamefont {Acosta}, \citenamefont
  {Doherty}, \citenamefont {Fatemi}, \citenamefont {Ivanov}, \citenamefont
  {Budker},\ and\ \citenamefont {Malinovsky}}]{Jarmola2020Robust_dup}%
  \BibitemOpen
  \bibfield  {author} {\bibinfo {author} {\bibfnamefont {A.}~\bibnamefont
  {Jarmola}}, \bibinfo {author} {\bibfnamefont {I.}~\bibnamefont {Fescenko}},
  \bibinfo {author} {\bibfnamefont {V.~M.}\ \bibnamefont {Acosta}}, \bibinfo
  {author} {\bibfnamefont {M.~W.}\ \bibnamefont {Doherty}}, \bibinfo {author}
  {\bibfnamefont {F.~K.}\ \bibnamefont {Fatemi}}, \bibinfo {author}
  {\bibfnamefont {T.}~\bibnamefont {Ivanov}}, \bibinfo {author} {\bibfnamefont
  {D.}~\bibnamefont {Budker}}, \ and\ \bibinfo {author} {\bibfnamefont {V.~S.}\
  \bibnamefont {Malinovsky}},\ }\href {\doibase
  10.1103/PhysRevResearch.2.023094} {\bibfield  {journal} {\bibinfo  {journal}
  {Phys. Rev. Research}\ }\textbf {\bibinfo {volume} {2}},\ \bibinfo {pages}
  {023094} (\bibinfo {year} {2020})}\BibitemShut {NoStop}%
\end{thebibliography}
\end{document}